**Phase-sensitive evidence for pair density wave in a kagome superconductor**

**Authors:** Xiao-Yu Yan[1]*, Guowei Liu[1]*, Hanbin Deng[1]*, Xitong Xu[2]*, Haiyang Ma[3]*, Hailang Qin[3]†, Junyi Zhang[4], Yuanyuan Zhao[3], Xiuhao Fan[1], Wei Song[1], Muwei Gao[1], Haitian Zhao[2], Zhe Qu[2], Yigui Zhong[5], Kozo Okazaki[5], Xiquan Zheng[6], Yingying Peng[6], Zurab Guguchia[7], Xianxin Wu[8], Da Wang[9], Qiang-Hua Wang[9], Hendrik Hohmann[10], Matteo Dürrnagel[10,11], Ronny Thomale[10], Jia-Xin Yin[1,3]†

**Affiliations:**

[1]State Key Laboratory of Quantum Functional Materials, Department of Physics, and Guangdong Basic Research Center of Excellence for Quantum Science, Southern University of Science and Technology, Shenzhen 518055, China.

[2]Anhui Key Laboratory of Low-Energy Quantum Materials and Devices, High Magnetic Field Laboratory of Chinese Academy of Sciences, Hefei 230031, Anhui, China.

[3]Quantum Science Center of Guangdong-Hong Kong-Macao Greater Bay Area, Shenzhen, China.

[4]William H. Miller III Department of Physics and Astronomy, Johns Hopkins University, Baltimore, Maryland 21218, USA.

[5]Institute for Solid States Physics, The University of Tokyo, Kashiwa, Japan.

[6]International Center for Quantum Materials, School of Physics, Peking University, Beijing 100871, China.

[7]Laboratory for Muon Spin Spectroscopy, Paul Scherrer Institute, CH-5232, Villigen PSI, Switzerland.

[8]CAS Key Laboratory of Theoretical Physics, Institute of Theoretical Physics, Chinese Academy of Sciences, Beijing 100190, China.

[9]National Laboratory of Solid State Microstructures & School of Physics, Nanjing University, Nanjing 210093, China.

[10]Institute for Theoretical Physics and Astrophysics, University of Wurzburg, 97074 Wurzburg, Germany.

[11]Institute for Theoretical Physics, ETH Zürich, 8093 Zürich, Switzerland.

*These authors contributed equally to this work.

†Corresponding authors. E-mail: yinjx@sustech.edu.cn, qinhailang@quantumsc.cn

**This PDF file includes:**

Main Text

Figure Legends 1 to 5

**Competing interests:** The authors declare no competing interests.

**Abstract: Pair-density-wave (PDW) exhibits periodic amplitude and sign modulations of the superconducting order parameter. Such a pairing state has long been proposed to be highly sensitive to nonmagnetic scattering, but its experimental realization remains elusive. Here we discover a nonmagnetic PDW-breaking effect in a kagome superconductor, using designer atomic nonmagnetic impurities and high-precision scanning tunneling microscopy (STM) at a base temperature of 30mK. We detect 2×2 pair density modulations by Josephson STM with a superconducting tip and 2×2 pairing gap modulations by normal STM. We find that the pairing modulations in both cases are substantially suppressed upon doping the kagome lattice with dilute isovalent nonmagnetic impurities, whereas the charge order and uniform superconductivity remain robust. We further identify the correlation between atomic dopants and the local suppression of PDW. We attribute these findings to a nonmagnetic pair-breaking effect, arising from the phase modulation of PDW in the kagome *d*-orbital. Taken together with its signatures in other state-of-the-art spectroscopy and transport measurements linked by theory, our findings support the ground state of the kagome superconductor as a correlated topological phase with superconducting loop currents.**

**Significance Statement:** Pair-density waves are an unusual form of superconductivity in which the superconducting order varies periodically in space, but phase-sensitive evidence for their microscopic nature has remained limited. Here we use normal and Josephson scanning tunneling microscopy to show that dilute nonmagnetic impurities strongly suppress pair-density-wave modulations in a kagome superconductor, while charge order and uniform superconductivity remain robust. These results reveal a nonmagnetic pair-breaking effect and support the phase sign reversals of the pair-density-wave in the kagome lattice.

**Main Text:** Conventional superconductivity is robust against nonmagnetic scattering as dictated by the Anderson theorem, and the pair-breaking effect induced by nonmagnetic impurities often signals the presence of an unusual superconducting order. The PDW is a kind of unconventional superconducting order (1), where the Cooper pairs carry finite momentum. A distinguishing feature of the intrinsic pair density wave (primary or induced) is the spatial phase sign modulation of the superconducting order parameter (Fig. 1*A*). Such a pairing state is then gapless (the gap in opened at the Fermi level when the Fermi momenta can be linked by the PDW vector, and other parts are left as Bogoliubov Fermi surfaces), and has long been proposed to be highly sensitive to nonmagnetic impurity scattering (2-6). It has recently been recognized that pair-breaking scattering interference is an alternative mechanism for pairing modulations (7). However, such modulations do not feature phase sign reversal and can be promoted with increasing impurity scattering. Therefore, the nonmagnetic impurity effect can serve as phase-sensitive evidence for PDW, which has been critically lacking in experiments owing to systematic challenges in finding a suitable platform and impurities and in performing high-precision measurements.

Recent measurements have gradually uncovered that the ground state of kagome superconductors $AV_3Sb_5$ (A = K, Rb, Cs) can host intrinsic PDW with possible phase sign modulations (8, 9). The ultra-low temperature STM (8-11), thermal transport (12, 13), muon spin rotation (14) (μSR), and nuclear quadrupole resonance (15) (NQR) reveal gapless excitations; normal and Josephson STM (8, 9, 16) reveal ubiquitous 2×2 pairing modulations in all $AV_3Sb_5$; scanning tunneling microscopy further reveals (8, 9) orbital selectivity that the uniform pairing mainly occurs in the Sb *p*-orbital, and kagome V *d*-orbital features residual Fermi arcs in connection with the pairing modulations; device-based chiral superconducting transport (17) and magnetic field tunability of the chirality of pairing modulations (8) further support time-reversal symmetry-breaking. These experimental progresses advance the theoretical understanding of the intertwined kagome orders (8, 18-29). Particularly, the 2×2 PDW can be self-consistently obtained (27) by considering attractive on-bond pairing near the *p*-type van Hove filling, which is chiral with broken time-reversal symmetry, exhibits gapless excitation with Bogoliubov Fermi arcs, and features charge 4e or 6e excitations that have been detected in device-based transport measurements (30). In addition to explaining these key unusual observations, the theory predicts a highly anisotropic gap in momentum space that has recently been confirmed by angle-resolved photoemission spectroscopy (ARPES) (31), the giant anomalous thermal Hall effect that has been confirmed by recent experiments (32), and a phase sign modulation feature that is yet to be verified. This progress highlights that the PDW in kagome superconductors can be probed by various state-of-the-art experimental techniques well beyond STM (illustrated in Fig. 1*B*).

Moreover, the theoretically envisioned chiral 2×2 PDW as the ground state of kagome superconductors features superconducting loop currents (27), in analogy with the Haldane model (33) for predicting the quantum anomalous Hall effect and the Varma model (34) for understanding the pseudogap phase in cuprates, as illustrated in (Fig. 1*C*). It deeply connects with the unconventional features of the charge order in the normal state (35-45). Such a loop current phase is driven by the frustrated pairing instability, and exhibits topological features such as chirality and anomalous thermal Hall effect in the superconducting state. Therefore, the PDW in kagome superconductors presents an excellent playground in the frontiers of the experimental and theoretical condensed matter physics. Yet, the most spectacular feature of the kagome PDW—the quantum phase sign reversal— is not uncovered.

In this work, we provide the key missing spectroscopic phase-sensitive evidence for the PDW in the kagome superconductor. We select the kagome superconductor $KV_3Sb_5$ as the clean platform (46), and we find that the nonmagnetic isovalent dopant Ta can be successfully doped into the V kagome lattice of $KV_3Sb_5$ single crystals (Fig. 2*A*) up to 4%, grown with a self-flux method (47, 48). Through susceptibility and transport measurements in Fig. 2*B*, we observe that the charge order is robust (slightly suppressed) while the superconductivity is substantially enhanced.

We then conduct dilution-refrigerator-based normal and Josephson scanning tunneling microscopy experiments on these crystals at a lattice temperature of 30mK and electronic temperature estimated as 90mK (Fig. S1). The underlying Ta dopant can be individually resolved in a high-bias-voltage image (49) of the Sb surface (Fig. 2*C*), which tightly bonds with the V kagome lattice. The concentration of the randomly distributed bright spots in the image matches the nominal Ta doping value of 4%. The zoomed-in image of each bright spot reveals that it is centered between two surface Sb atoms, aligning with the underlying V atomic position in the kagome lattice. The 2×2 charge order can be revealed by the low-bias (50mV, an energy close to the charge order gap) topographic image in Fig. 2*D*. A comparison of the topographic data with the pristine sample reveals that the 2×2 charge order is robust (Fig. 2*E* and Fig. S4), consistent with the robust charge order detected by susceptibility. The robustness of the charge order is also confirmed by our X-ray measurements of the superlattice peaks in Fig. S6.

Moreover, we find that the tunneling spectrum at the superconducting energy scale alters dramatically, with the pairing gap substantially increased in the doped crystal (Fig. 2*F*), consistent with the enhanced superconductivity detected by transport.

The pairing condensate of the crystal is further supported by our tunneling experiment using a superconducting Nb tip, revealing a much larger total pairing gap (Fig. 2*G*). We detect a zero-bias peak in the differential conductance by reducing the tip–sample distance and the characteristic double kink features in the related current–voltage spectrum, both of which serve as key signatures of the Josephson tunneling signal and indicate the Cooper pair tunneling between two superconductors (8, 10, 50-52). We have noticed the increasing conductance between the Josephson peak and coherence peaks, which has been repeatedly seen in the Josephson tunneling between kagome superconductors and a Nb tip (8, 10). It is likely related to the residual Fermi states in the kagome superconductor and remains an open question. We confirm that there is no noticeable change in the sample area after taking the spectra after reducing the tip-sample distance.

The scanned Josephson tunneling signal can detect the spatial modulation of the phenomenologically defined Cooper pair density (8, 52) and provide evidence for PDW. Particularly, for superconductors with a smaller energy gap, the pair density can be effectively detected (8, 52) by $N_{\mathrm{J}}(\boldsymbol{r}) \propto g_{\mathrm{J}}(\boldsymbol{r}, E=0) \times R_{\mathrm{N}}^{2}(\boldsymbol{r})$, where $g_{\mathrm{J}}(\boldsymbol{r}, E = 0)$ is the spatially resolved zero-bias peak in the differential conductance data at position $r$ and $R_{\mathrm{N}}(r)$ is the spatially resolved Josephson junction normal state resistance at a bias where the current-voltage characteristics are linear. The scanned pair density $N_{\mathrm{J}}(\boldsymbol{r})$ data on the Sb surface with atomic resolution is shown in Fig. 2*H*. The corresponding Fourier transform data $N_{\mathrm{J}}(\boldsymbol{q})$ in right panel of Fig. 2*H* demonstrates the existence of 2×2 PDW in the material, which is expected from the intertwining between 2×2 charge order and superconductivity.

A straightforward interplay of the slightly weakened charge order $\rho_Q$ and substantially enhanced uniform superconductivity $\Delta_0$ predicts an enhanced PDW order $\Delta_{\mathrm{PDW}} \propto \rho_Q \Delta_0$ by Ginzburg–Landau density functional analysis (53), an intriguing phenomenon to explore. Unexpectedly, a comparison of this 2×2 pair density vector peak measured under similar conditions with Josephson STM for pristine (8) and doped samples in Fig. 3*A* and Fig. S4 reveals that the global 2×2 pair density modulation is substantially suppressed by 65%. Note that we have rotationally symmetrized $N_{\mathrm{J}}(\boldsymbol{q})$ each data for comparison, and the conclusion is similar for unsymmetrized data for all three $\boldsymbol{Q}_{2\times2}$ directions that we have checked. Even the strongest 2×2 vector peak for the doped sample is weaker than the weakest 2×2 vector peak for the pristine sample, thus we omitted the chiral anisotropy for this comparison study. The strong PDW suppression is in stark contrast with the robust charge order and enhanced superconductivity.

To validate this striking observation of PDW suppression, we measure the modulations of the pairing gap with a normal tip. We measure the low-energy differential conductance $g(r,E)$ along a line shown in Fig. 3*B*. Through precise measurements of the superconducting coherence peaks, we detect tiny modulations in their energy positions (see Fig. S2 for further elaboration on the spatial-energy resolution). We extract the energy positions of the coherence peaks to obtain the modulation of the pairing gap $\Delta(r)$ as shown in the right panel of Fig. 3*B*, which is compared with the pristine sample (8). It is evident that the *2a* pairing gap modulation is also substantially suppressed by Ta dopants. To quantify this finding, we measure the gap modulations in the two-dimensional gap map, and the 2×2 modulation of the pairing gap in absolute value is revealed by their Fourier transformation analysis in Fig. 3*C* and Fig. S4, demonstrating that the global 2×2 pairing gap modulation is suppressed by 66%, which is remarkably consistent with the 2×2 pair density suppression. The pairing modulation at the Bragg vector is widely observed in experiments on other superconductors (8, 10, 51-57). Its invariance against Ta dopants in absolute value suggests a trivial origin of the modulation at the Bragg vector (without sign reversals). We note that the $N_J(\boldsymbol{r})$ is measured in the intensity channel of the tunneling signal with the superconducting tip, while $\Delta(\boldsymbol{r})$ is measured differently and independently in the energy channel with a normal tip, and their quantitative consistence on the suppression of the 2×2 modulations highlight the robustness of our finding.

Therefore, both the pair density modulation and pairing gap modulation reveal a strongly suppressed PDW, as illustrated using the pair density data $N_J(\boldsymbol{q})$ in Fig. 4*A* and the corresponding filtered $N_J(\boldsymbol{r})$ data in Fig. 4*B*. Before we discuss the theoretically envisioned nonmagnetic PDW suppression effect, we need to rule out other reasons, including an artifact of STM measurement as the set-point effect, dopant-induced strong Fermi surface change, and magnetic scattering from the dopant. Firstly, we measure the zero-energy differential conductance map with a normal tip with the same junction setup as the gap map. Then, we find that the intensities of the related 2×2 vector peak are comparable for pristine and doped data in the upper panel of Fig. 4*C* (see also Fig. S4), in line with the aforementioned robust charge order by susceptibility and X-ray measurements. Thus, the observed pairing modulation suppression cannot arise from a simple set-point effect. Secondly, our first-principles calculations demonstrate that the 4% Ta dopants do not change much of the Fermiology and the low-energy electronic structure in Fig. S8, which is consistent with the isovalent nature of Ta dopant and is supported by the ARPES measured Fermi surfaces in pristine and doped samples in Fig. S7. Thirdly, we rule out strong magnetic scattering as the origin. In the lower panel of Fig. 4*C*, we show that the local superconducting gap is rather robust at the Ta dopant site. It suggests the nonmagnetic nature of the Ta dopants, which is supported by our first-principles calculations in Fig. S9 that Ta is unlikely to be magnetic or to induce magnetism. Thus, based on our systematic transport, susceptibility, STM, ARPES, X-ray, and first-principles calculations, we rule out other reasons and conclude that the strong PDW suppression is more likely due to the nonmagnetic PDW-breaking effect.

To further support the PDW breaking effect, we investigate the spatial correlation between PDW and atomic dopants. At the same atomic position, we obtain a topographic image at high bias voltage to reveal the locations of the Ta dopants and the low-energy superconducting gap map in Fig. 4*D*. The Fourier transform of the gap map reveals the 2×2 PDW order (see Fig. 4*E* inset). An inverse Fourier transform of the 2×2 vector peaks in the gap map extracts the 2×2 gap modulation (Fig. 4*E*), revealing spatial inhomogeneity of PDW. Larger local 2×2 gap modulations correspond to the larger local PDW order. We then extract the local amplitude of the 2×2 pairing gap modulation (58), which defines the local PDW strength, as shown in Fig. 4*F*. Electronic inhomogeneity is often associated with unconventional ground states and correlated with dopant's position (59). Therefore, we mark the previously determined atomic dopant position on the PDW map in Fig. 4*F*. We quantified the spatial correlation between the Ta dopant positions and the local PDW strength and obtained a cross-correlation coefficient of -0.25, indicating a moderate anticorrelation. This result suggests a statistical tendency for the PDW signal to be weaker near Ta dopants. The magnitude of the anticorrelation may be limited by the interference of the impurity effects, impacts from deeper dopants, and our finite spatial-energy resolution.

Lastly, we discuss the mechanism and implications of the nonmagnetic PDW-breaking effect. The unique lattice geometry of the kagome lattice lies in the fact that it features three sublattices, shown in Fig. 5*A*. At the *p*-type van Hove filling as relevant to the $KV_3Sb_5$ system, the Fermi surface is hexagonal like as shown in Fig. 5*B*, where the van Hove singularities are located at M points featuring large density of states. The Bloch wave function at each M point is contributed by one sublattice. The sublattice interferences can promote the formation of PDW (22-27). Our calculation of pairing susceptibility (27) in Fig. 5*C* reveals strong instability at $Q_{2\times2}$, corresponding to the vector connecting M points in Fig. 5*B*. Owing to the sublattice interference, the resulting 2×2 PDW, as shown in Fig. 5*D*, carries chiral orbital currents (27). Such a PDW ground state explains the essential experimental observations, including gapless excitations, 2×2 pairing modulations, residual Fermi arcs, time-reversal symmetry breaking, and charge 6e thermal excitations, as we mentioned in the introduction. In addition, although the calculated real space gap modulations are as tiny as that in the experiment (a few percentage of the total pairing gap) owing to the frustrated lattice geometry, it predicts (27) a large gap anisotropy in momentum space (Fig. 5*E*) and a giant anomalous thermal Hall effect. The recent ARPES measurements confirm the predicted large gap anisotropy in the kagome *d*-orbital (31), as marked in Fig. 5*E*, and the giant anomalous thermal Hall effect in the superconducting state has also recently been detected (32).

The theory also predicts spatial phase sign reversals of the PDW state (27) as illustrated by the small arrows in the inset of Fig. 5*D*. It has long been proposed that such PDW order with phase sign reversals is highly sensitive to nonmagnetic impurity scattering (2-4). Nonmagnetic impurities cause scattering between states with different momenta and phases, thereby tending to destroy the corresponding spatial modulations (2-4). This physics is confirmed by our study of the nonmagnetic PDW-breaking effect using the Abrikosov-Gor'kov approach (60-62) based on our PDW model, as shown in Fig. 5*F* (See Supplemental Material for more details). The PDW order is progressively suppressed by increasing the nonmagnetic scattering strength. Thus, we attribute the striking PDW suppression to the nonmagnetic PDW-breaking effect, originating from the phase modulation of the PDW order in the kagome *d*-orbital. We note that the phase here refers to the phase of the PDW order parameter, rather than the phase difference between the tip and sample in a Josephson junction formed with a superconducting tip.

The nonmagnetic pair-breaking phenomenon uncovered here is a fundamental quantum effect of PDW, highlighting PDW as a distinct pairing state. Such an effect provides phase-sensitive evidence for PDW and we look forward to using this effect to test the proposed PDW in other quantum systems. Taken together with other detected quantum effects in kagome superconductors, including the chiral pairing modulations (8, 9, 16), gapless excitation (8-15) with *d*-orbital Bogoliubov Fermi arcs (8, 9), orbital-selective anisotropic pairing (31), charge 4e or 6e thermal excitations (30), and superconducting diode effect (17) and giant thermal Hall effect (32), our finding suggests that the kagome PDW as the ground state of kagome superconductors embraces both correlation and topology, effectively realizing the condensed Haldane model with superconducting loop currents. The research of kagome PDW also presents an example where state-of-the-art measurements and advanced theory progressively validate each other to promote our understanding of correlated topological phase and to uncover new quantum effects. The systematics developed in this field can be useful for exploring other emergent quantum matter.

## Figures

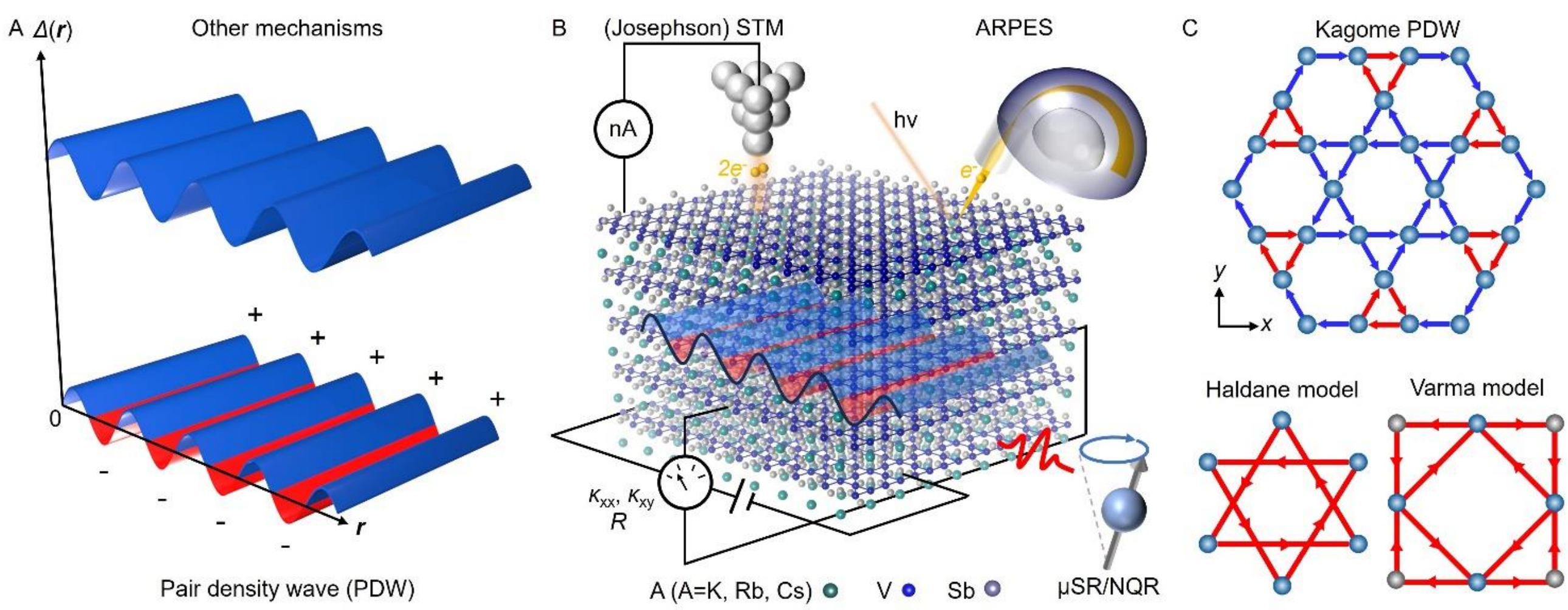

**Fig. 1.** PDW physics and its detection in the kagome superconductor. (*A*) Comparison between PDW and pairing modulations produced by other mechanisms. The PDW uniquely features a phase sign modulation in real space. The real-space modulation periodicity is $2\pi/\mathbf{Q}$, which characterizes the finite momentum pairing between electrons with momenta $-\boldsymbol{k}$ and $+\boldsymbol{k}+\mathbf{Q}$. (*B*) Schematic illustration of PDW in kagome superconductors $AV_3Sb_5$ (A = K, Rb, Cs), which can be probed by various state-of-the-art techniques, including (Josephson) STM, ARPES, device-based electrical and thermal transport, μSR and NQR. (*C*) The upper panel illustrates the self-consistently calculated $2\times2$ chiral PDW in the kagome lattice (27), where the arrows denote the superconducting loop currents. The lower panels illustrate the Haldane model (33) in the honeycomb lattice for predicting the quantum anomalous Hall effect and the Varma model (34) for understanding the pseudogap phase in cuprates. Such an analogy highlights the kagome PDW as a correlated topological phase in condensed matter.

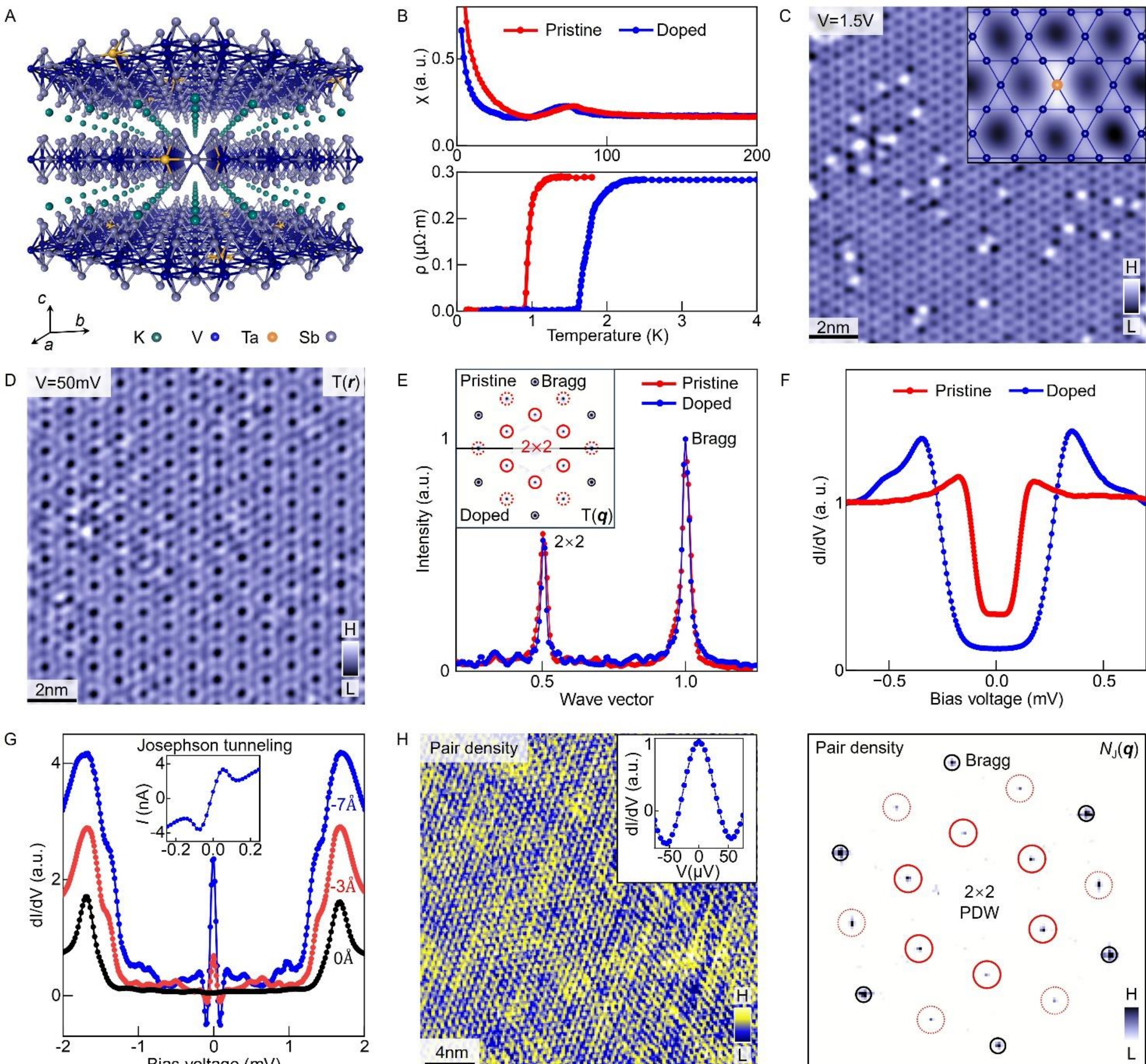

**Fig. 2.** Resolving intertwined electronic orders in a kagome superconductor. (*A*) Crystal structure of 4% Ta-doped $KV_3Sb_5$. (*B*) The upper panel compares the susceptibility data on charge order transition. The lower panel compares the resistivity data on superconducting transition. (*C*) Topographic image of the Sb surface under a high bias voltage, showing individual Ta dopants as bright spots whose counting is 4%. The inset shows the zoomed-in image of a bright spot and the indication of the atomic details, showing its consistency with a Ta dopant in the underlying kagome lattice. (*D*) Topographic image of the Sb surface under a low bias voltage, showing the 2×2 charge order. (*E*) The inset compares the Fourier transform of topography for a larger area, revealing 2×2 charge order, marked by red circles. Black circles indicate Bragg peaks, and the red dashed circles highlight higher-order peaks. The main panel compares the corresponding vector peaks, where both data are rotationally symmetrized from inset. (*F*) Comparison of the superconducting gap for pristine and doped samples. (*G*) Differential conductance taken with reducing the tip-sample distance from 0Å to -7Å in reference to normal condition using a superconducting Nb tip, showing Josephson signal at zero bias. The inset shows a typical *I*–*V* curve with double kink features. (*H*) Pair density map (left, inset shows the Josephson zero-bias peak signal) and its Fourier transform (right), exhibiting 2×2 PDW as marked by the red circles.

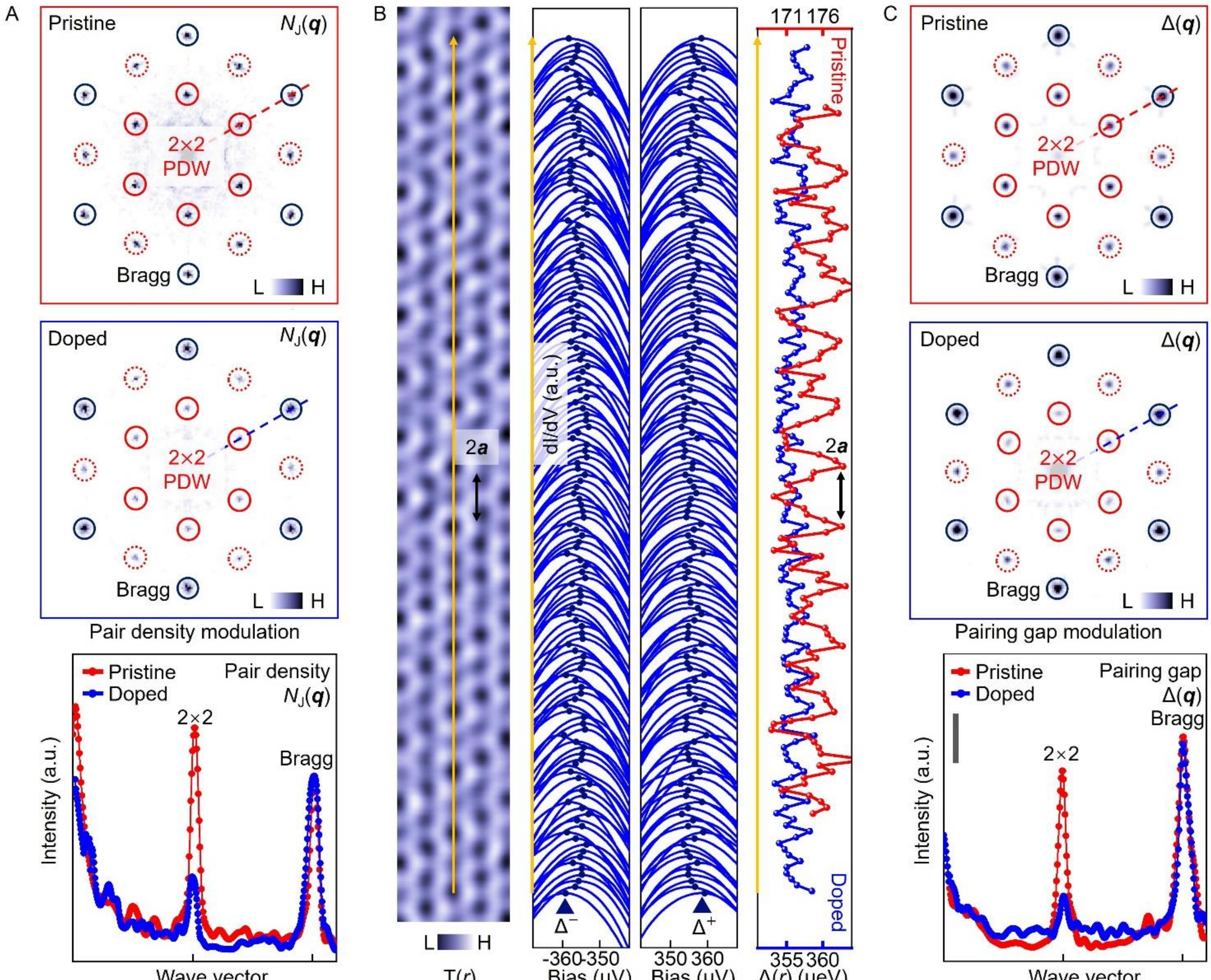

**Fig. 3.** Strong suppression of the PDW detected by Josephson and normal STM. (*A*) Comparison of the pair density modulation data for the pristine (upper panel) and doped (middle panel) samples in the ***q***-space. Both data are rotationally symmetrized. Their intensities along the dashed lines are compared in the lower panel, showing a 65% reduction of the 2×2 pair density modulation. (*B*) High-resolution spectra (middle) near the superconducting gap edge, along the line drawn in the topographic image (left) obtained with a normal tip. $\Delta^{\pm}(r)$ is determined from the energy position of the peak, as indicated by the dark blue dots. The right panel shows the extracted pairing gap modulations $\Delta(r) = [\Delta^{+}(r) - \Delta^{-}(r)]/2$, which is further compared with the gap modulations of the pristine sample (energy scale noted in upper axis with same energy segregation ratio). (*C*) Comparison of the pairing gap modulation data for the pristine (upper panel) and doped (middle panel) samples in the *q*-space. Both data are rotationally symmetrized. Their intensities along the dashed lines are compared in the lower panel, showing a 66% reduction of the 2×2 pairing gap modulation. The vertical bar denotes a value whose inverse Fourier transform corresponds to a modulation amplitude of 2μeV (from peak to bottom).

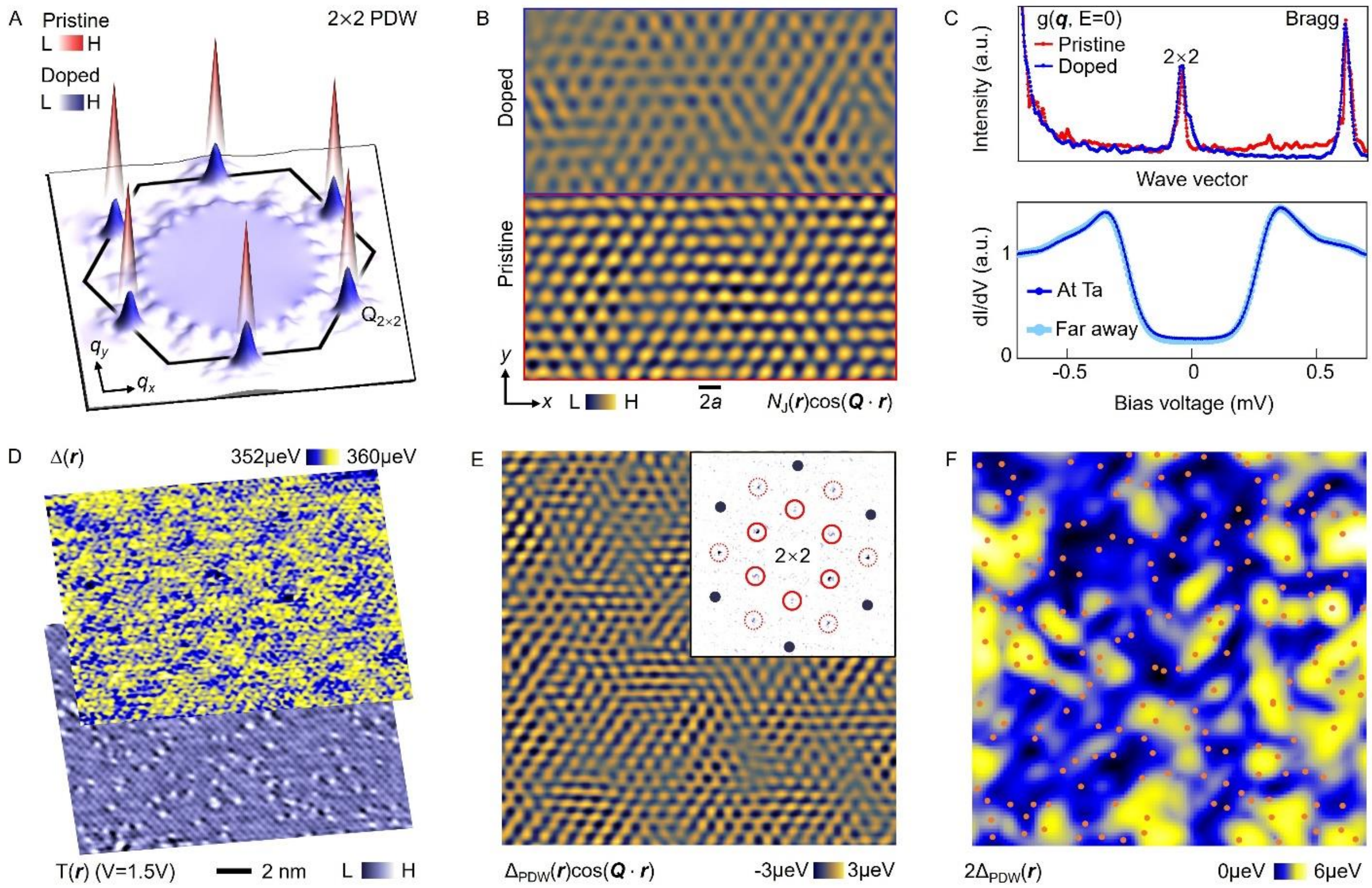

**Fig. 4.** Suppression of the PDW order by dopants. (*A*) Illustration of the observed PDW suppression in doped samples. The 2×2 vector peaks in pair density modulations are strongly suppressed by nonmagnetic isovalent Ta impurities in the kagome lattice. (*B*) Inverse Fourier transform of the 2×2 vector peaks for the pair density map, showing PDW suppression in real space. (*C*) Upper panel compares the symmetrized Fourier transform data of the zero-energy differential conductance maps measured with the same junction setup as that for pairing maps. The lower panel compares the superconducting gap at the dopant and distant positions. (*D*) Lower figure: High bias topography. Upper figure: Pairing gap map at the same position. (*E*) Inverse Fourier transform of the 2×2 vector peaks from the gap map, showing 2×2 gap modulations. The inset displays its Fourier transform, showing the 2×2 PDW order. (*F*) PDW map obtained by extracting the local amplitude of the 2×2 gap modulations. The dots indicate dopants.

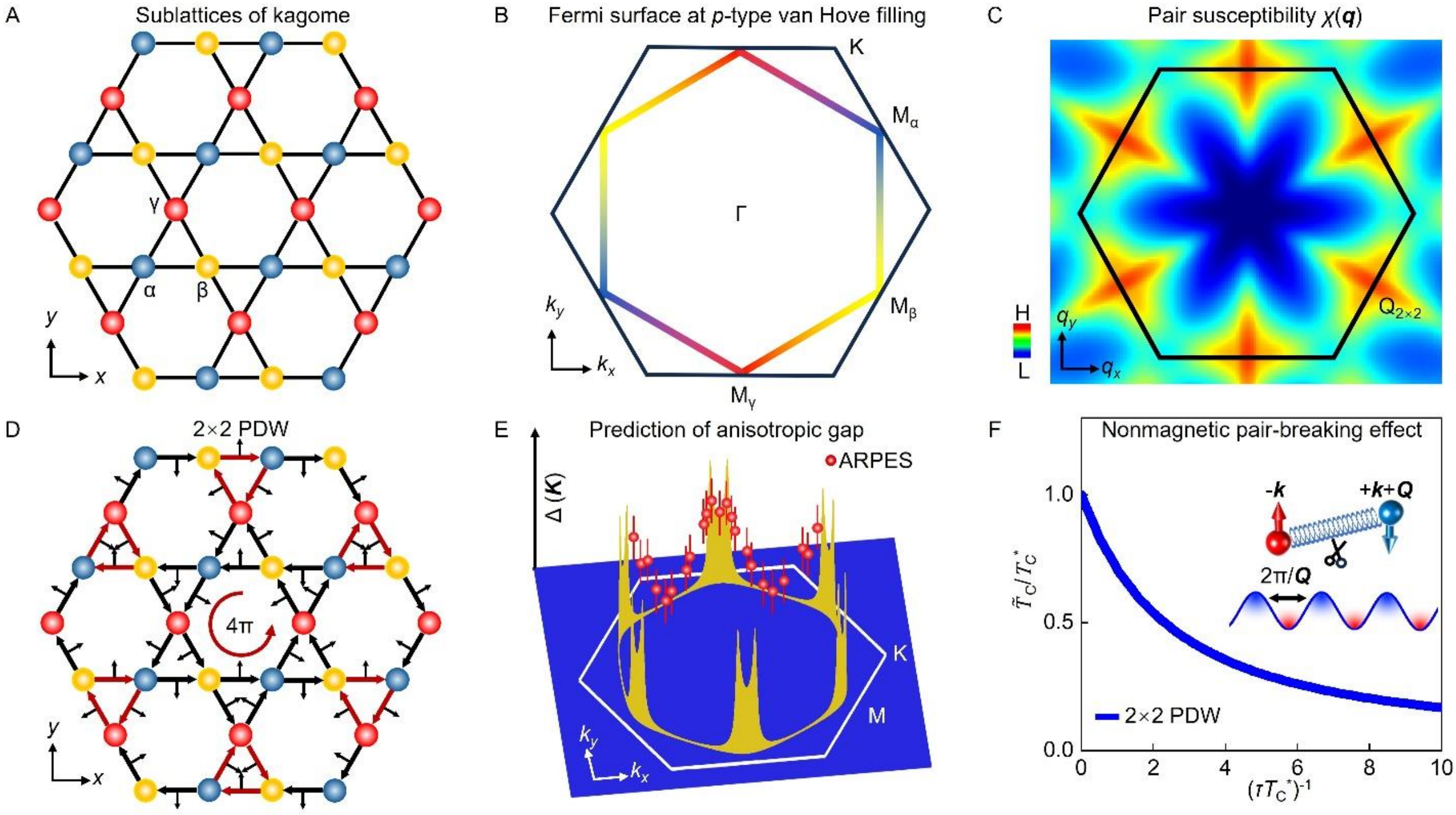

**Fig. 5.** Mechanism of nonmagnetic PDW-breaking effect. (*A*) Unique geometry of the kagome lattice: it features three sublattices denoted as α, β and γ with different colors. (*B*) Fermi surface at *p*-type van Hove filling of the kagome lattice. The color denotes the wave functions from the three sublattices. (*C*) Calculated pairing susceptibility showing pairing instabilities at $Q_{2\times2}$. (*D*) Calculated 2×2 PDW pattern. The thick arrows denote the orbital currents, and the thin arrows denote the different phase angles. (*E*) Predicted gap anisotropy in comparison with ARPES data (red spheres). (*F*) Calculated suppression of the PDW order parameter by increasing the nonmagnetic scattering strength. The inset illustrates the nonmagnetic PDW-breaking effect owing to the spatial phase sign modulations of the PDW.

**Acknowledgments:** We thank Sofie Castro-Holbæk, Manfred Sigrist, Hong Yao, Fuchun Zhang, Congjun Wu, Hu Miao, S. S. Islam, Yang Gao, Sen Zhou, Andreas Kreisel, and Brian Møller Andersen for discussions. J.Z. acknowledges the encouragement and support from Yi Li. We acknowledge the support from the National Key R&D Program of China (grant numbers 2023YFA1407300 and 2025YFA1411500), the National Natural Science Foundation of China (Grant numbers 12374060, 12474153, and 11804163), Guangdong Provincial Quantum Science Strategic Initiative (Grant No. GDZX2401001 and grant No. GDZX2501002), the Young Scientists Fund of National Natural Science Foundation of China (funding number: 12504162), and the Guangdong Natural Science Foundation (Grant No. 2026A1515010666). H.Q. acknowledges the start-up fund (funding number: QD2301007) from Quantum Science Center of Guangdong-Hong Kong-Macao Greater Bay Area. H.H., M.D., and R.T. were supported by the Deutsche Forschungsgemeinschaft (DFG, German Research Foundation) through the Würzburg-Dresden Cluster of Excellence ctd.qmat – Complexity, Topology and Dynamics in Quantum Matter (EXC 2147, project-id 390858490). M.D. is furthermore grateful for a PhD scholarship of the Studienstiftung des Deutschen Volkes. X.X. was supported by the CAS Project for Young Scientists in Basic Research (Grant No. YSBR-084), the CAS Project (Grant No. JZHKYPT-2021-08), Anhui Provincial Major S&T Project (s202305a12020005), Anhui Provincial Natural Science Foundation (No. 2508085ZD013), and the High Magnetic Field Laboratory of Anhui Province (contract No. AHHM-FX-2020-02). X.W. acknowledges support from the National Natural Science Foundation of China (Grant Nos. 12574151, 12447103, and 12447101). Z.G. acknowledges support from the Swiss National Science Foundation (SNSF) through SNSF Starting Grant (No. TMSGI2 211750).

**Supplementary Information for**
**"Phase-sensitive evidence for pair density wave in a kagome superconductor"**

**Authors:** Xiao-Yu Yan[1]*, Guowei Liu[1]*, Hanbin Deng[1]*, Xitong Xu[2]*, Haiyang Ma[3]*, Hailang Qin[3]†, JunYi Zhang[4], Yuanyuan Zhao[3], Xiuhao Fan[1], Wei Song[1], Muwei Gao[1], Haitian Zhao[2], Zhe Qu[2], Yigui Zhong[5], Kozo Okazaki[5], Xiquan Zheng[6], Yingying Peng[6], Zurab Guguchia[7], Xianxin Wu[8], Da Wang[9], Qiang-Hua Wang[9], Hendrik Hohmann[10], Matteo Dürrnagel[10,11], Ronny Thomale[10], Jia-Xin Yin[1,3]†

**Single crystal growth**

Single crystals of Ta-doped $KV_3Sb_5$ were grown using the self-flux method from the constituent elements. High-purity K, V, Ta, and Sb were mixed in an alumina crucible in a molar ratio of 5:2.7:0.3:13 and sealed in an evacuated quartz ampoule. The ampoule was heated to 1273K, soaked for 20*h*, then cooled down to 923K at 2K/*h*. Residual flux was removed by centrifugation. The chemical composition of the as-grown single crystals was determined to be $K(V_{0.96}Ta_{0.04})_3Sb_5$ via energy-dispersive X-ray spectroscopy, and checked with the topographic data in scanning tunneling microscopy. Single crystal X-ray diffraction measurements were conducted using the custom-designed X-ray instrument at 20K. The lattice parameters for $KV_3Sb_5$ are determined to be $a = b = 5.464$Å, $c = 8.909$Å, while for 4% Ta-doped $KV_3Sb_5$, the lattice parameters are $a = b = 5.464$Å, $c = 8.899$Å.

**Normal and Josephson scanning tunneling microscopy**

Over 40 crystals were cleaved and imaged in this project, where most cleaved crystals are used to explore the suitable conditions for tunneling at various voltages for different purposes. Crystals with dimensions up to 2mm×2mm×0.5mm are cleaved mechanically *in-situ* at 77K under ultra-high vacuum conditions and immediately inserted into the microscope head, which is pre-cooled to the $He^4$ base temperature of 4.2K. The microscope head is then further cooled to 30mK using a dilution refrigerator. The cooling procedure takes about 12*h*. Tunneling conductance spectra are obtained with Ir/Pt tips (or a superconducting tip, as described below) using standard lock-in amplifier techniques. Each crystal is extensively scanned to identify large, clean Sb surfaces, which typically takes a week to find. Topographic images are typically obtained with a tunneling junction set to $V = 50$mV, $I = 0.5$nA. To image the individual Ta dopants, the topographic images are taken with a tunneling junction set to $V = 1.5$V, $I = 0.5$nA. Conductance and gap maps are acquired by taking a spectrum at each location (with the feedback loop off) with a tunneling junction set of: $V = 1$mV, $I = 1$nA, and modulation voltage $V_m = 3\sim20$μV. These tunneling conditions further require a clean sample area and stable, atomic-resolution tunneling tips. One method of preparing a stable tip is to anneal them on an Au crystal under similar tunneling conditions, ensuring the tip is robust.

We discuss the spatial energy resolution of scanning tunneling microscopy. To estimate the electronic temperature of our system, we measure the superconducting gap of a related full-gap kagome superconductor (14%-Ta doped $CsV_3Sb_5$) with similar setup. By fitting the superconducting gap to the BCS gap function (Fig. S1), we estimate the electronic temperature to be approximately 90mK. The commonly discussed energy resolution, $\Delta E$, can be understood as follows: if an electronic state exhibits δ function, then it will behave as a Gaussian-like function in the tunneling data, with a full width at half maximum of $dE$ (Fig. S2*A*). In our case, $dE = 90\text{mK}\cdot3.5k_B$, where $k_B$ is the Boltzmann constant.

At a fixed position, if two such states have an energy separation smaller than $dE$, they will merge into a single peak in the tunneling data, making it impossible to distinguish the two states directly (Fig. S2*B*). However, if these states are located at different locations, their spectral peak energy difference, although broadened, may still be distinguished by scanning tunneling spectroscopy (Fig. S2*C*). In this case, both the spatial resolution of the tunneling tip and the signal-to-noise ratio of the tunneling spectrum are as crucial as the electronic energy resolution $dE$, which can be improved by maintaining a stable tunneling condition and elongating the measuring time. In the present case, we find that scanning tunneling microscopy can resolve the spatial modulations of superconducting coherence peaks at μeV level. For similar reasons, previous studies (1-4) achieve 1~10μeV level spatial energy resolution in studying PDW even at a lattice temperature of 300~400mK. In principle, the higher the electronic temperature, the flatter the measured peak, and a higher signal-to-noise ratio is required. And if the sample's coherence peak is too broad, then a superconducting tip can help to enhance the sharpness of the measured coherence peak, enhancing the signal-to-noise ratio for determining the peak energy (or enhance the effective "energy resolution"). In our case, the sample's coherence peak is narrower than that of Nb, and with a Nb tip, the measured coherence peak is actually broader, which does not contribute to an effective higher energy resolution in determining the PDW gap modulations. This is also the reason we measure the pairing gap map with normal tip.

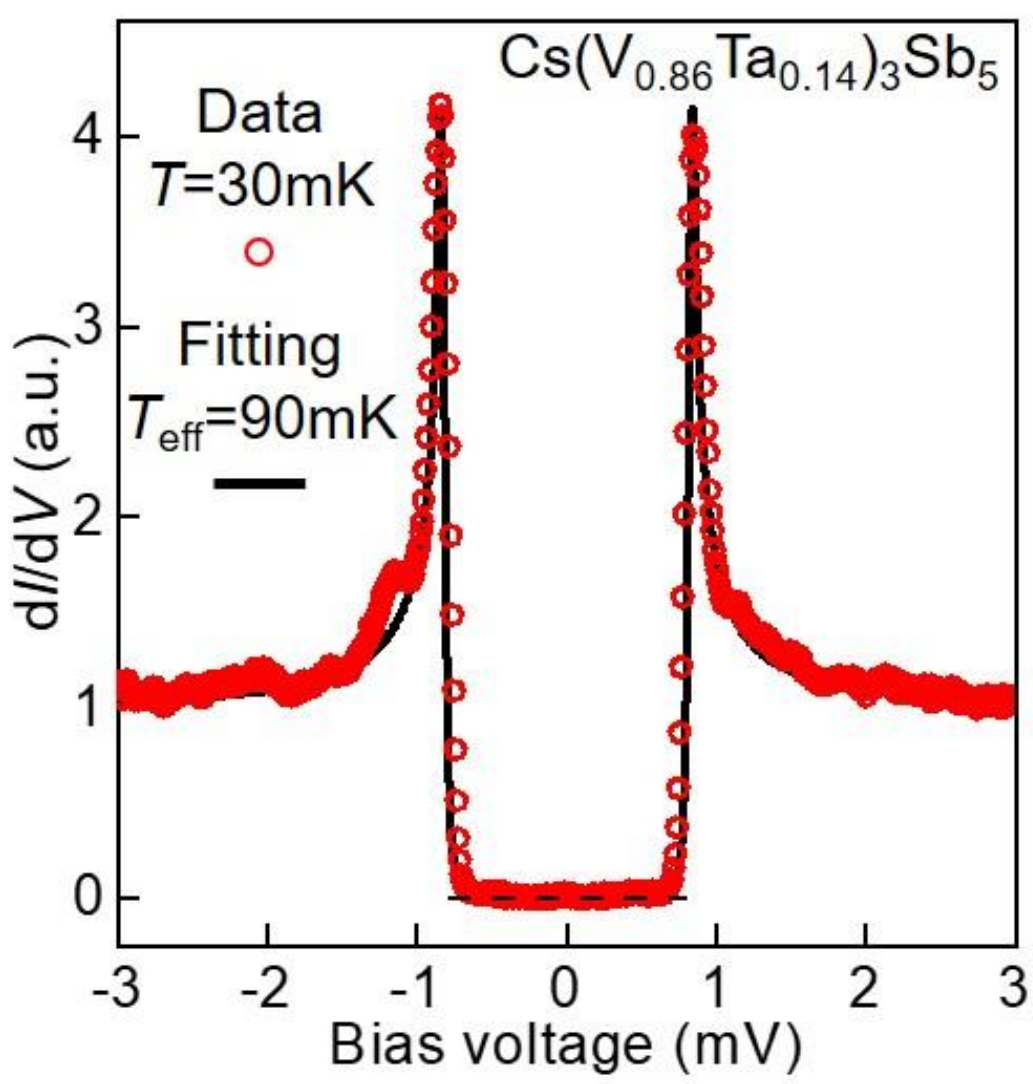

**Fig. S1. Estimation of electronic temperature.** Fitting the superconducting gap of $Cs(V_{0.86}Ta_{0.14})_3Sb_5$ single crystal with the BCS gap function ($V$ = 1mV, $I$ = 1nA, $V_m$ = 20μV). The fit estimates the electronic temperature of our system to be 90mK.

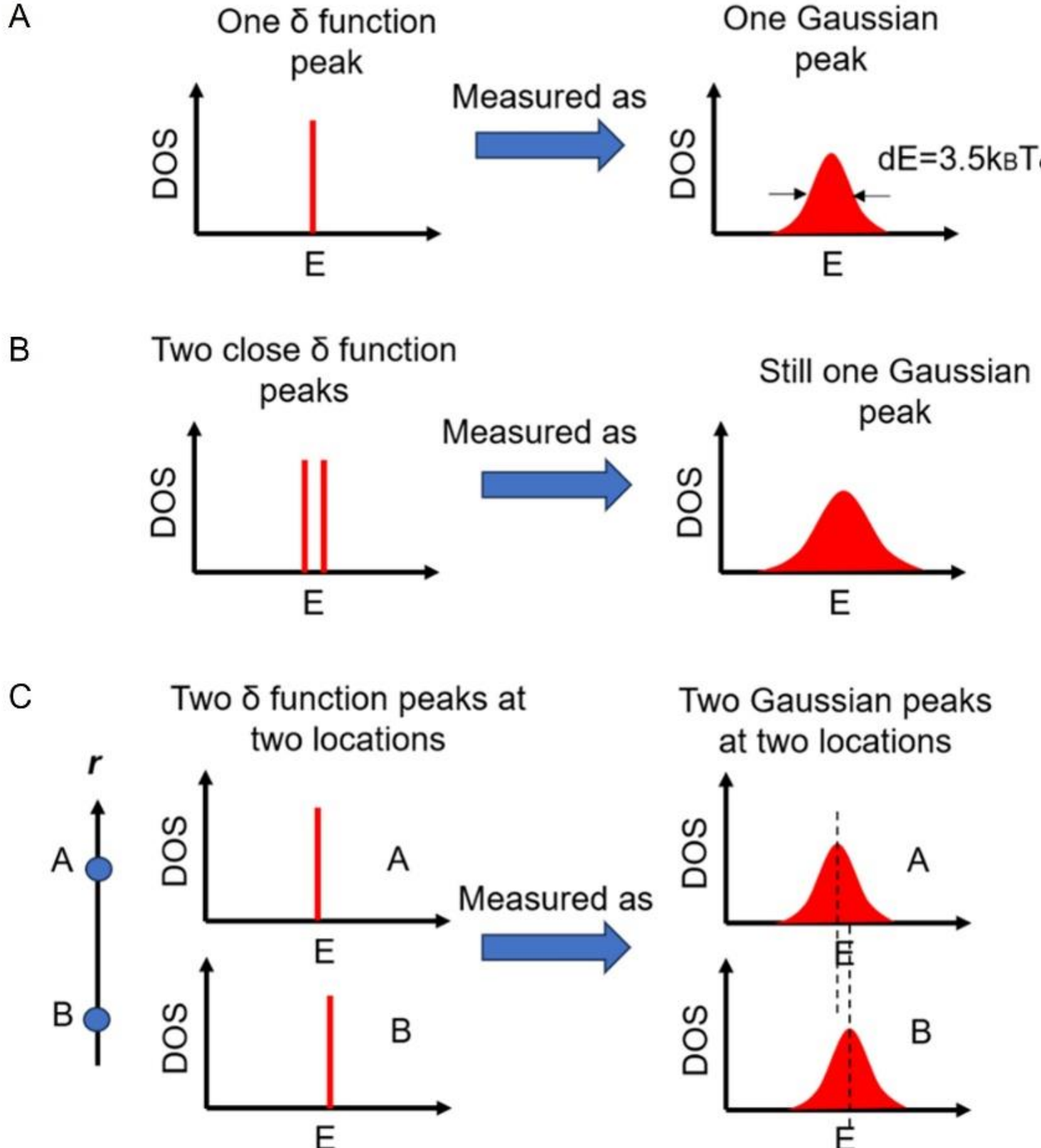

**Fig. S2. Spatial energy resolution of scanning tunneling microscopy.** (*A*) A δ function state is measured as a Gaussian-like peak in the tunneling spectrum due to the convolution effect with a derivative of the Fermi-Dirac distribution function. The Gaussian-like peak has a finite width of $dE = 3.5k_BT_{eff}$, where $k_B$ is the Boltzmann constant and $T_{\mathrm{eff}}$ is the effective electronic temperature of the system. (*B*) If two close δ function states exist, they will be measured as a single Gaussian-like peak when their energy separation is smaller than *dE*. In this case, tunneling spectroscopy cannot distinguish the two states due to finite electronic temperatures. (*C*) If two close δ function states are located differently at real space, they can be measured as two Gaussian-like peaks (one at each location). Their peak energy difference may be smaller than *dE*, depending on the measurement accuracy. To resolve both peaks, the following conditions are required: i) high spatial resolution to prevent peak merging, and ii) high signal-to-noise ratio to accurately determine the peak energy positions. For similar reasons, previous studies (1-4) can achieve μeV level spatial energy resolution in studying PDW even at a lattice temperature of ~300mK.

Superconducting polycrystalline Nb tips were used for the Josephson scanning tunneling microscopy experiment. In order to prepare a superconducting Nb tip, we use electrochemical etching of Nb wires, followed by e-beam heating. More specifically, we use an NaOH solution with a concentration of ~52 wt%, and etch the tip with an AC voltage of around 30~50V between the Nb wire (with a diameter of 0.3mm) and carbon electrode for about 20 minutes, while the solution is maintained at about 90℃. After the etching, the tip is typically rinsed in de-ionized water before loading into the ultra-high vacuum chamber. The tip is then heated to an estimated temperature of 1500℃ for about 10 seconds for ten times with electron-beam heating for further cleaning. To check the tip superconducting gap size and stability, we typically first test on a polycrystalline Pb sample by topographic imaging and taking d*I*/d*V* spectrums.

The pairing gap of the Nb tip is estimated to be 1.33meV based on the total gap size of the Josephson junction. Due to the complex surface environment (Sb surface mixed with K surfaces, adatoms, and other disordered surfaces), it is extremely challenging to maintain the superconductivity (pairing gap) of the tip with atomic resolution. We find that when a large bias voltage of 1.5V is applied to resolve the Ta dopant, the tip often crushes into the sample and loses its superconductivity properties. Therefore, topographic images are usually taken with the tunneling junction set to $V$ = 50mV, $I$ = 0.5nA. Josephson conductance spectra and maps are obtained by measuring the spectrum at each location (with the feedback loop off) using a tunneling junction setup of: $V$ = 3mV, $I$ = 8nA, and modulation voltage $V_m$ = 30μV. The zero-bias peak is readily observed at this junction setup and is further enhanced by increasing the tunneling current. The Josephson coupling energy, $E_J$, is estimated to be 20mK·$k_B$ (where $k_B$ is the Boltzmann constant), which is smaller than the electronic temperature of the system 90mK, placing the Josephson junction close to the phase diffusive limit. In principle (5), the phenomenologically defined local pair density $N_J$(r) is expected to be proportional to the Josephson critical current $I_J(\boldsymbol{r})$ times the normal state resistance $R_N$(r): $N_\mathrm{J}(\boldsymbol{r}) \propto I_\mathrm{J}^2(\boldsymbol{r})R_\mathrm{N}^2(\boldsymbol{r})$. However, for small gap superconductors where the Josephson coupling energy $E_J$ is typically smaller than the electronic temperature, accurately measuring $I_J$ is challenging. In such cases, the tunneling current takes the form: $I = 0.5I_\mathrm{J}^2 ZV/(V^2 + V_C^2)$, where Z is the high-frequency impedance and $V_C$ is the characteristic voltage at which the maximum phase-diffusive electron-pair tunneling current occurs. Then the differential conductance at zero energy is given by $\mathrm{g_J}(0) = dI/\mathrm{dV} = 0.5I_\mathrm{J}^2 V_\mathrm{C}^{-2} \propto I_\mathrm{J}^2$. Thus, $\mathrm{g_J}(\boldsymbol{r}, E = 0)R_\mathrm{N}^2(\boldsymbol{r})$ provides a practical method for measuring the pair density $N_\mathrm{J}(\boldsymbol{r})$ in small gap superconductors. Therefore, we measure both zero-bias conductance peak $\mathrm{g_J}(\boldsymbol{r}, E = 0)$ and normal-state junction resistance $R_\mathrm{N}(\boldsymbol{r})$ at each location $\boldsymbol{r}$. The normal-state resistance $R_\mathrm{N}(\boldsymbol{r})$ is determined as $V/I(r, V)$, where $V$ is the sample bias at which the current-voltage characteristics are linear (typically at 3mV) and $I(r, V)$ is the tunneling current at bias $V$.

To further validate this phenomenological scaling in our measurements, we examined the dependence of the measured zero-bias conductance $g(0)$ on $1/R_\mathrm{N}^2$ under different junction setups. As shown in Fig. S3, both pristine and Ta-doped $KV_3Sb_5$ exhibit an approximately linear relationship between $g(0)$ and $1/R_\mathrm{N}^2$, consistent with the expectation for phase-diffusive Josephson tunneling.

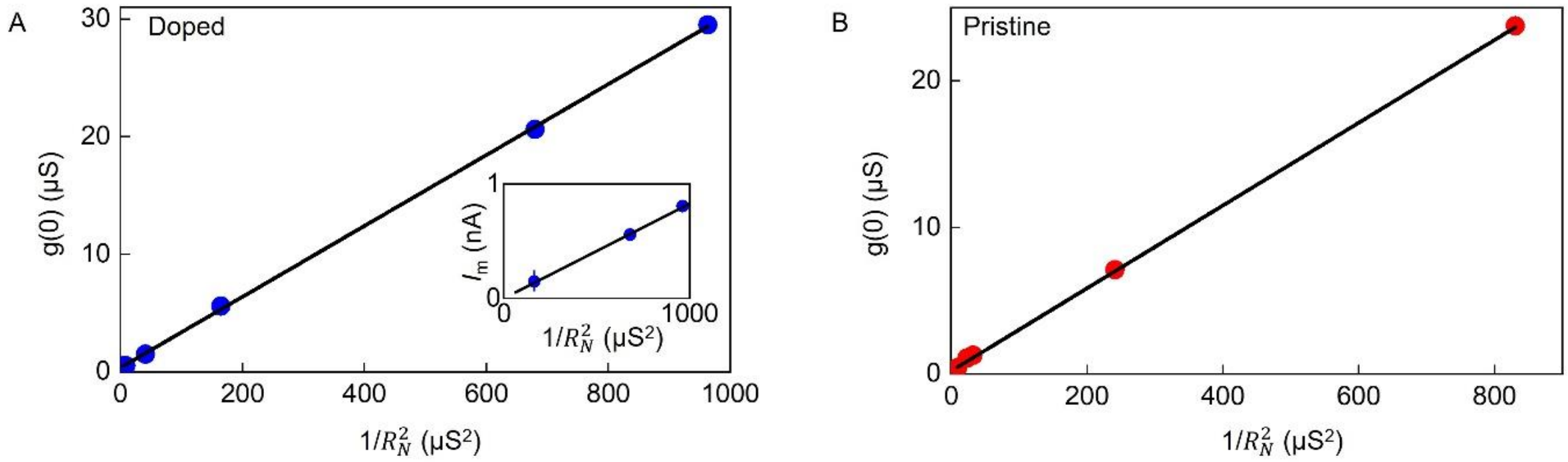

**Fig. S3. Scaling relation for Josephson tunneling.** Measured $g(0)$ versus $1/R_{\mathrm{N}}^2$ at different junction setups for (A) pristine and (B) Ta-doped $KV_3Sb_5$, showing an approximately linear scaling in both samples. The inset in (B) shows the measured $I_m$ versus $1/R_{\mathrm{N}}^2$ for Ta-doped $KV_3Sb_5$, which is also approximately linear.

**Extended Data for charge order and PDW comparison with pristine sample**

Fig. S4 shows the related topography for charge order ($V$ = 50mV, $I$ = 0.5nA), pair density maps ($V$ = 3mV, $I$ = 8nA, $V_m$ = 30μV), pairing gap maps ($V$ = 1mV, $I$ = 1nA, $V_m$ = 3μV) and zero-energy quasi-particle interference maps (V = 1mV, I = 1nA, $V_m$ = 50μV) for both pristine and doped crystals, as the extended data for Figs. 2, 3 and 4 in the main text. Apparent low-frequency noise has been filtered for better clarity.

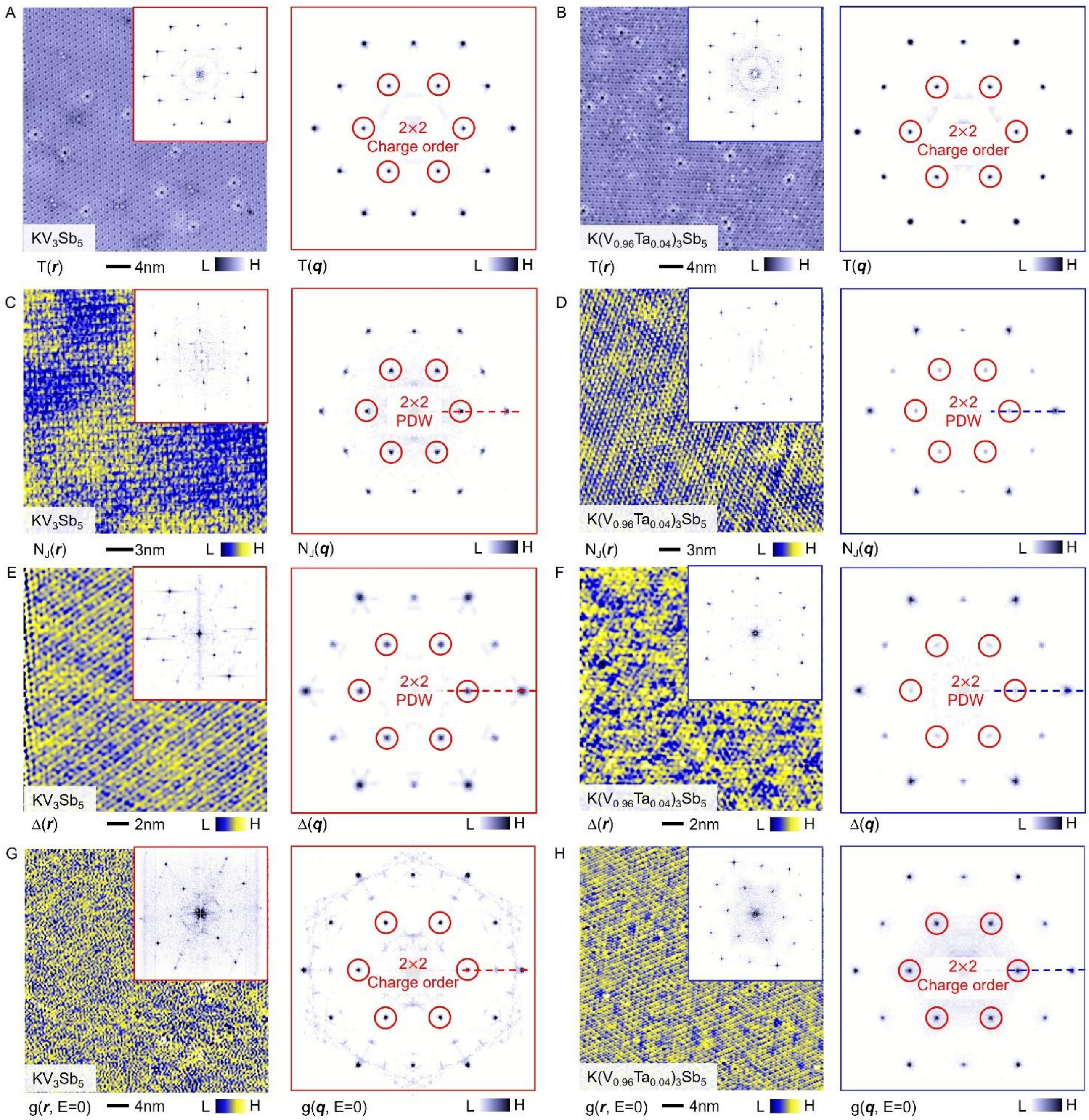

**Fig. S4. Charge order and PDW comparison with pristine sample.** (*A*, *B*) The left data shows topography for measuring charge order for pristine and doped crystals, respectively. Their insets show the related Fourier transform. The right panel plots the related Fourier transform with six-fold symmetrization, showing 2×2 charge order. (*C*, *D*) The left data shows pair density map measured with a Nb tip for pristine and doped crystals, respectively. Their insets show the related Fourier transform. The right panel plots the related Fourier transform with six-fold symmetrization. (*E*, *F*) The left data shows the gap map measured with a normal tip for pristine and doped crystals, respectively. Their insets show the related Fourier transform. The right panel plots the related Fourier transform with six-fold symmetrization. (*G*, *H*) The left data shows the zero-energy map measured with a normal tip for pristine and doped crystals, respectively. The right panel plots the related Fourier transform with six-fold symmetrization.

Fig. S5 shows another set of pair density map $N_J(r)$ taken under different junction setup conditions, $V = 1$ mV and $I = 1$ nA for 4% Ta-doped $KV_3Sb_5$ and $V = 2$ mV and $I = 3$ nA for $KV_3Sb_5$, respectively, compared with those discussed in the main text (Fig. 3, $V = 3$ mV, $I = 8$ nA for both). As can be seen, the peak intensities in Ta-doped $KV_3Sb_5$ are strongly suppressed compared to those from pristine $KV_3Sb_5$.

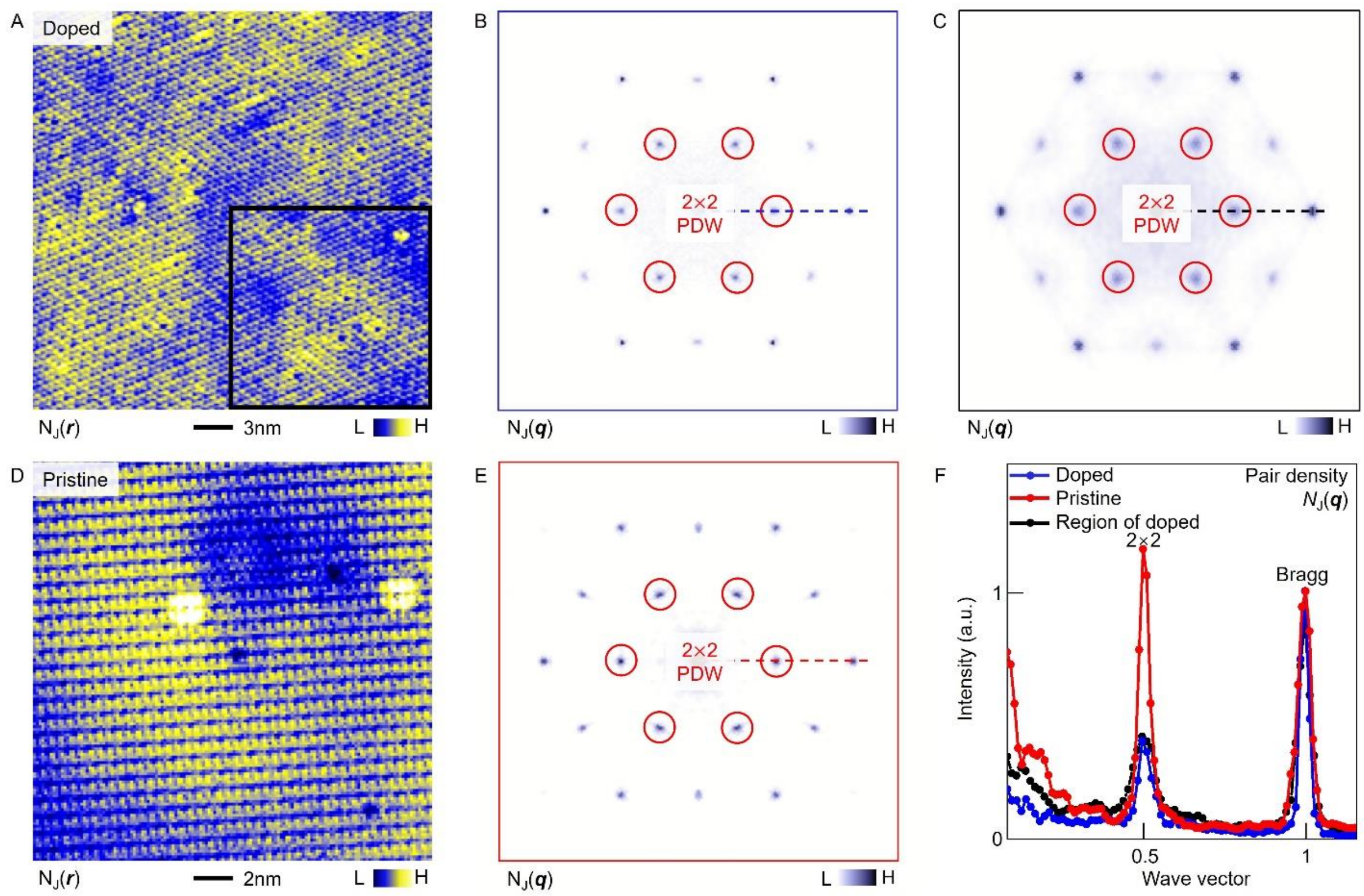

**Fig. S5. Robustness of the PDW suppression under different tunneling conditions and scan regions.** (*A*) Pair density map $N_J(r)$ measured on Ta-doped $KV_3Sb_5$ with a Nb tip under a different tunneling condition from that used in the main text. The black square marks a selected subregion used for additional analysis. (*B*) Fourier transform of the full $N_J(r)$ map in (A), showing the $2 \times 2$ PDW peaks. (*C*) Fourier transform of the black-boxed subregion in (A). (*D*) Pair density map $N_J(r)$ measured on pristine $KV_3Sb_5$ with a Nb tip under another tunneling condition. (*E*) Fourier transform of the $N_J(r)$ map in (*D*), showing the corresponding $2 \times 2$ PDW peaks in pristine $KV_3Sb_5$. (*F*) Comparison of the linecuts taken along the dashed directions in (*B*), (*C*), and (*E*). The $2 \times 2$ PDW peak in Ta-doped $KV_3Sb_5$ remains clearly suppressed relative to that in pristine $KV_3Sb_5$, demonstrating that the main conclusion of PDW suppression is robust against variations in tunneling condition and scan size.

**Extraction of the local amplitude of the 2×2 pairing gap modulation and cross-correlation analysis**

The pairing gap modulation in the real-space image consists of a series of modulation wavevectors (3, 5-7): $\Delta(\mathbf{r}) = \sum_{\boldsymbol{Q}} \Delta_{\boldsymbol{Q}}(\boldsymbol{r}) e^{-i\boldsymbol{Q}\cdot\boldsymbol{r}}$, where $\Delta_{\boldsymbol{Q}}(\boldsymbol{r})$ is the complex amplitude at wavevector $\boldsymbol{Q}$ and position $\boldsymbol{r}$; $e$ is the Euler's number and $i$ is the imaginary unit. The Fourier transform of this real-space image can be given by: $\Delta(\boldsymbol{q}) = F[\Delta(\boldsymbol{r})] = \sum_{\boldsymbol{Q}} \int \mathrm{d}\boldsymbol{q}'\, \Delta_{\boldsymbol{Q}}(\boldsymbol{q}')\delta(\boldsymbol{Q} - \boldsymbol{q} + \boldsymbol{q}') = \sum_{\boldsymbol{Q}} \Delta_{\boldsymbol{Q}}(\boldsymbol{q} - \boldsymbol{Q})$, where $\Delta_{\boldsymbol{Q}}(\boldsymbol{q} - \boldsymbol{Q})$ is the Fourier transform of the complex amplitude centered at $\boldsymbol{Q}$. $\Delta_{\boldsymbol{Q}}(\boldsymbol{q})$ can be extracted by shifting it back to the center associated with wavevector $\boldsymbol{Q}$. The inverse Fourier transform of $\Delta(\boldsymbol{q})$ can be written as: $\Delta_{\boldsymbol{Q}}(\boldsymbol{r}) = F^{-1}[\Delta_{\boldsymbol{Q}}(\boldsymbol{q})] = F^{-1}\left[F[\Delta(\boldsymbol{r})e^{i\boldsymbol{Q}\cdot\boldsymbol{r}}] \cdot \frac{1}{\sqrt{2\pi}\sigma_{\boldsymbol{q}}} e^{-\frac{\boldsymbol{q}^2}{2\sigma_{\boldsymbol{q}}^2}}\right]$, where $\sigma_{\boldsymbol{q}}$ is the cutoff length, which is the inverse of the averaging length scale in real space, but larger than the filter length at $\boldsymbol{Q}$. We collected the gap map $\Delta_{\mathrm{SC}}(\boldsymbol{r})$ as the real-space image $\Delta(\mathbf{r})$. We obtained the $\Delta_{\mathrm{SC}}(\boldsymbol{q})$ in *q*-space by applying a Fourier transform to the $\Delta_{\mathrm{SC}}(\boldsymbol{r})$, where the $\boldsymbol{Q}_{\boldsymbol{i}}(\boldsymbol{i} = 1,2,3)$ wavevectors at 2×2 peaks reveal the PDW order. Using the two-dimensional lock-in method, we extract the local PDW strength in real space associated with tentative modulation wavevector $\boldsymbol{Q}_{\boldsymbol{i}}(\boldsymbol{i} = 1,2,3)$: $A_{\mathrm{PDW}}(\boldsymbol{r}) = \sum_{\boldsymbol{i}} |\Delta_{\mathrm{PDW}}^{\boldsymbol{Q}_i}(\boldsymbol{r})| = \sum_{\boldsymbol{i}} |F^{-1}[\Delta_{\mathrm{SC}}^{\boldsymbol{Q}_i}(\boldsymbol{q})]| = \sum_{\boldsymbol{i}} |F^{-1}\left[F[\Delta_{\mathrm{SC}}(\boldsymbol{r})e^{i\boldsymbol{Q}_i\cdot\boldsymbol{r}}] \cdot \frac{1}{\sqrt{2\pi}\sigma_{\boldsymbol{q}}} e^{-\frac{\boldsymbol{q}^2}{2\sigma_{\boldsymbol{q}}^2}}\right]|$. We then quantified the spatial correlation between the Ta dopant positions and the local PDW strength and obtained a cross-correlation coefficient of -0.25, indicating a moderate anticorrelation.

**X-ray evidence for robust charge order against dilute Ta dopant**

Single crystal X-ray diffraction measurements were conducted using the custom-designed X-ray instrument at 20K. The instrument is equipped with a Xenocs Genix3D Mo K$\alpha$ (17.48keV) x-ray source, delivering $2.5 \times 10^7$ photons/sec within a beam spot size of 150$\mu$m at the sample position. The samples were mounted on a Huber 4-circle diffractometer. Diffraction signals were collected using a highly sensitive single-photon counting PILATUS3 R 1M solid-state area, featuring a pixel array of 980 × 1042 pixels. Each pixel measures 172$\mu$m × 172$\mu$m. Three-dimensional mapping of momentum space was achieved by taking images in 0.1° increments while rotating samples. Fig. S6*A* displays the (L, H) map of reciprocal space at K = 0.5 and (H, K) map at L=11. 2×2×2 charge density wave peaks are observed. A comparison with the pristine sample under identical conditions is shown in Fig. S6*B*. The two samples exhibit similar charge density wave intensities, as shown in Fig. S6*C*. Further systematic analysis, incorporating all the measured charge density wave peaks and the method described in Ref. (8), confirms that the 4% Ta dopants have minimal impact on the 2×2×2 charge density wave structure, as shown in Fig. S6*D*. The lattice parameters for $KV_3Sb_5$ are determined to be a=b=5.464Å, c=8.909Å, while for 4% Ta-doped $KV_3Sb_5$, the lattice parameters are a=b=5.464Å, c=8.899Å.

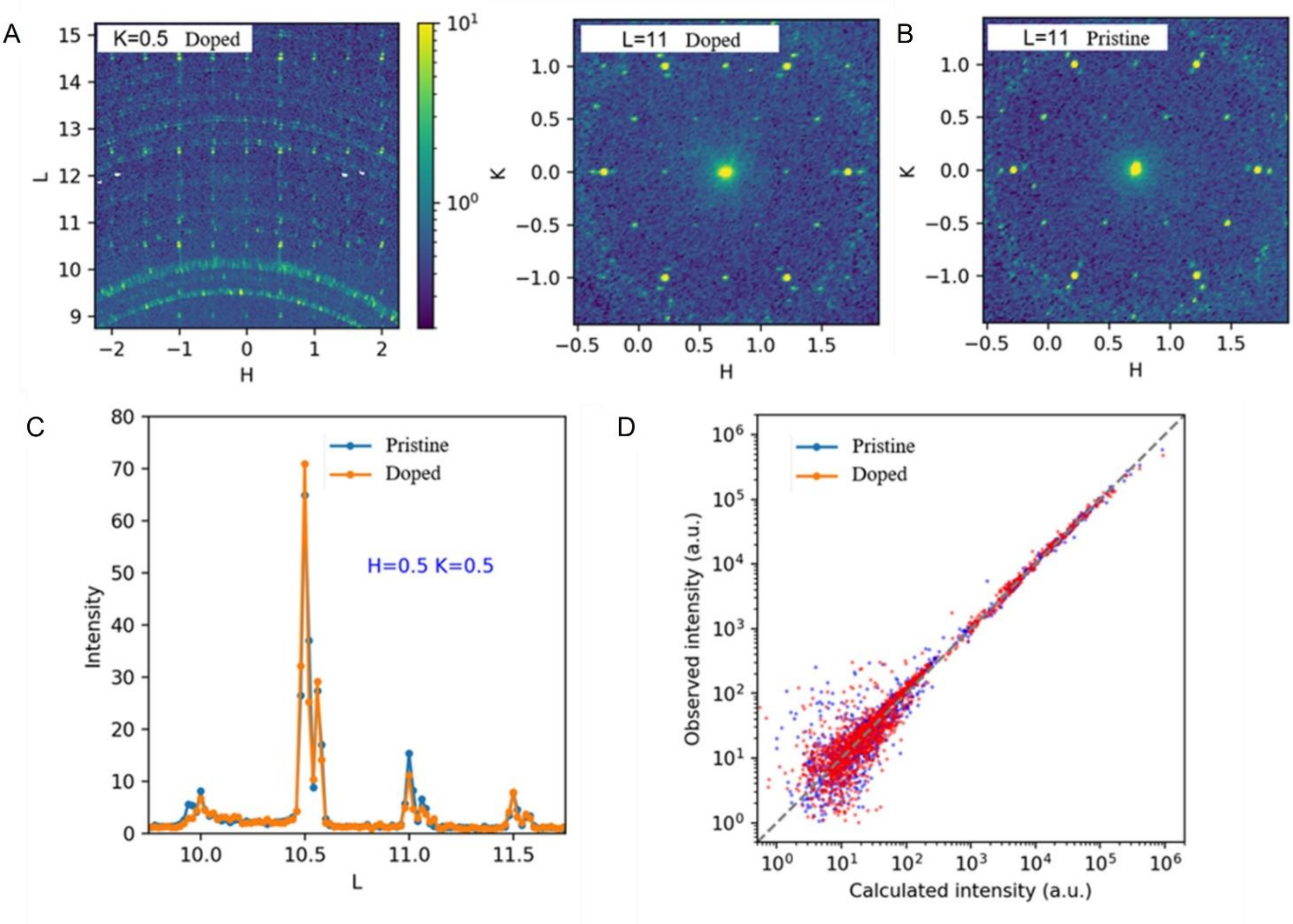

**Fig. S6. X-ray diffraction measurement confirming the robust charge order.** (*A*) (L, H) map of the reciprocal space at K = 0.5 and (H, K) map at L=11 for 4% Ta-doped sample. (*B*) (H, K) map of reciprocal space at L=11 for the pristine sample. (*C*) Comparison of the charge density wave profile between the two samples in the L direction through the (0.5,0.5,11) peaks, showing identical intensities. The double peaks arise from the small non-monochromaticity of the Mo x-ray source, which consists of $K_{\alpha 1}$ (17.48keV) and $K_{\alpha 2}$ (17.37keV). (*D*) Observed X-ray diffraction intensities from all peaks compared to the calculated 2×2×2 structures, showing that both samples exhibit identical charge density wave-induced lattice displacements.

**Robust Fermiology against dilute Ta dopant**

To examine the effects of 4%-Ta doping on the Fermi surface topology, we performed ARPES measurements on both pristine and 4%-Ta-doped $KV_3Sb_5$. For the ARPES measurements, we used a photon energy of 21.218eV obtained from a Helium discharge lamp (Scienta Omicron VUV 5050) and detected the photoelectrons using a Scienta R4000 analyzer. The single crystals were cleaved *in-situ* and maintained in a vacuum of better than $3\times10^{-11}$ torr throughout the measuring process. Fig. S7 shows Fermi surface mappings at $T$ = 120K, where a direct comparison between pristine and 4%-Ta-doped $KV_3Sb_5$ reveals no detectable differences in the Fermi surface topology. To explore this further, we overlaid the Fermi surface contours derived from the ARPES studies (9) on $KV_3Sb_5$ onto our intensity maps. These contours, scaled to best fit our data, align consistently for both pristine and the 4%-Ta-doped samples, suggesting that 4%-Ta doping has no detectable effect on the Fermi surface size, consistent with its isovalent nature.

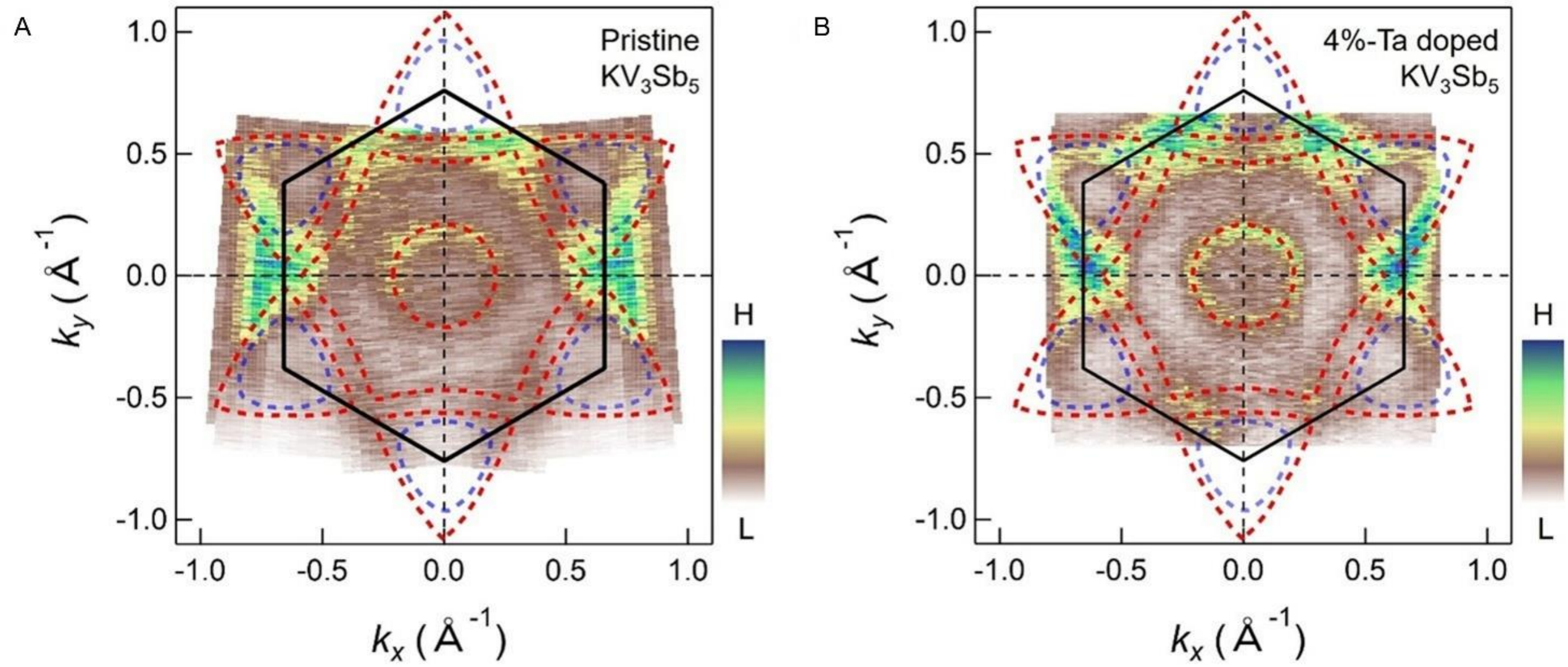

**Fig. S7. ARPES measurements.** (*A*, *B*) Fermi surface maps of pristine and 4%-Ta-doped $KV_3Sb_5$, respectively. Data were obtained by integrating the photoemission intensity within ± 10meV from $E_F$ at $T = 120K$. The dashed lines represent Fermi surface contours derived from previous ARPES studies on $KV_3Sb_5$ in Ref. (9).

**First-principles calculations on the robust charge order and nonmagnetic nature of Ta impurity**

First-principles density-functional-theory (DFT) calculations are performed using the Vienna ab initio simulation package (VASP) which adopts the projector augmented-wave method (10). The energy cutoff is set at 400eV and exchange-correlation functional of the Perdew-Burke-Ernzerhof (PBE) type (11) is used for the electronic structure calculations. The doping effects of Ta are treated in virtual-crystal approximation (12) as implemented in VASP. Phonopy code (13) together with finite-difference method are used for the calculations of the phonon spectrums, a 3×3×2 supercell and a 5×5×3 k-mesh is adopted.

Using the virtual-crystal approximation, we are able to grasp the main features of the electronic structures of $KV_3Sb_5$ as a function of Ta doping. From Fig. S8*A*, we can see that at 4% doping of Ta, the overall low-energy electronic structures show no obvious changes. Under such Fermi surfaces, the charge order transition temperature can be roughly captured by the smearing temperature adopted during DFT calculations of the phonon spectrum. With increasing smearing $\sigma$, the soft phonon modes around the M and L points gradually becoming harder, as shown in Fig. S8*B*. The charge order transition temperature $T^*$ sets the onset where the soft mode disappears. By including the anharmonic corrections (14), we are able to track $T^*$~120K for the pristine sample, which is of the same order as that detected in experiments. We then calculate $T^*$ as a function of the doping, we find that doping 4% Ta to $KV_3Sb_5$ will reduce $T^*$ by 6%, which aligns with the experimentally observed robust charge order.

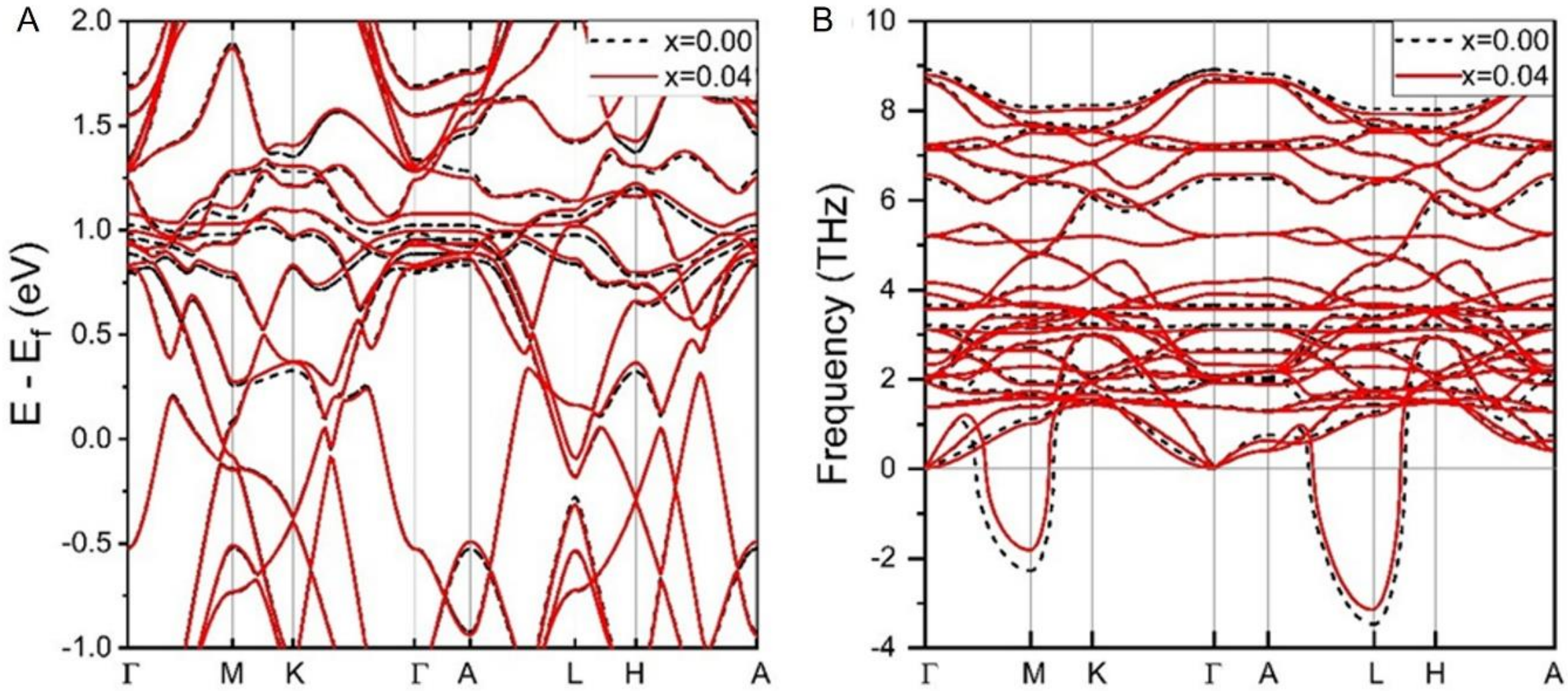

**Fig. S8. Calculated impact of Ta dopants on electronic and phonon structure.** (*A*) Calculated band structures for different doping concentrations of $K(V_{1-x}Ta_x)_3Sb_5$. (*B*) Calculated phonon spectrum for different doping concentrations of $K(V_{1-x}Ta_x)_3Sb_5$.

For calculations of the superconducting properties, we use the EPW code (15), which is interfaced with Quantum Espresso (16) and Wannier90 packages (17). The energy cutoff of the plane-wave basis set is set at 80Ry. The Gaussian smearing method with a width of 0.004 Ry and a 10×10×6 k-mesh was employed for obtaining the ground-state electronic structure. The lattices of the pristine and doped structure from VASP relaxation are used. A 10×10×6 k-mesh and 5×5×3 q-mesh were adopted as coarse grids and the 50×50×30 k-mesh and 25×25×15 q-mesh were utilized as dense grids. After obtaining the Eliashberg spectral function (18) $\alpha^2F(\omega)$, the superconducting transition temperature $T_c$ was calculated with the Allen-Dynes formula (19), $T_c = \frac{\omega_{log}}{1.2}\exp\left[-1.04\frac{1+\lambda}{\lambda-(0.62\lambda+1)\mu^*}\right]$, where $\lambda = 2\int\frac{d\omega}{\omega}\alpha^2F(\omega)$ is the total electron-phonon-coupling strength, and $\mu^*$ is the effective screened Coulomb repulsion constant, which is empirically (20, 21) set to 0.16 and $\omega_{log} = \exp\left[\frac{2}{\lambda}\int\frac{d\omega}{\omega}\alpha^2F(\omega)\ln(\omega)\right]$ is the logarithmic average of the Eliashberg spectral function. From Table S1 we can find that with doping 4% Ta to $KV_3Sb_5$, the superconducting transition temperature increases from 1.0K to 1.3K, which supports the enhanced superconductivity observed in experiments. The larger enhancement of superconductivity in experiments may be due to the competition of uniform superconductivity and PDW in the d-orbital channel, which is beyond our first-principles calculation and deserves further analysis and modeling.

| Doping | Tc (K) | $\lambda$ | $\omega_{log}$ | $\Delta$ (meV) | $\mu^*$ |
|---|---|---|---|---|---|
| 0.00 | 1.013013 | 0.6352795 | 6.482754 | 0.153639 | 0.16 |
| 0.04 | 1.296265 | 0.6747295 | 6.553155 | 0.196598 | 0.16 |

Table S1. Calculated parameters for the Allen-Dynes formula.

Lastly, we evaluate that whether the Ta dopant can induce magnetic scattering or magnetism in the system. The static generalized susceptibility of a system at wavevector $\boldsymbol{q}$ in the random-phase-approximation (RPA) is written as: $\hat{\chi}_{RPA}(\boldsymbol{q}) = \hat{\chi}^0(\boldsymbol{q}) \cdot [1 + U(\boldsymbol{q}) \cdot \hat{\chi}^0(\boldsymbol{q})]^{-1}$, where $\hat{\chi}^0(\boldsymbol{q})$ is the bare generalized susceptibility (22), $U(\boldsymbol{q})$ is the on-site Kanamori-Hubbard interaction matrix (22). The calculations are performed with Wannier tight-binding Hamiltonians, which are obtained by VASP2WANNIER90 interface (17). 60 orbitals are considered including the *p* orbitals of K and *d* orbitals of V or Ta and spin degree of freedom. To find the effects of Ta doping on the Fermi surface instability, we calculated the generalized RPA susceptibility of $K(V_{1-x}Ta_x)_3Sb_5$ at different doping concentrations, the results are shown in Fig. S9. For susceptibilities at the four high-symmetry points, G, M, K and A, only at A and G points show diverging behaviors when increasing the value of the on-site Hubbard *U*. At the A point the susceptibilities host the largest diverging modes (Fig. S9*A*). Through careful analysis, we find that there are 3 leading diverging modes at *U* = 3.0eV, which turn out to be the 3 ferromagnetic orders with local magnetic moments along the **x**, **y** and **z** axis, respectively, consistent with previous results (22). Moreover, with increasing the doping of Ta, we can see in Fig. S9*B* that the divergences are getting weaker, which means that the doping of Ta will suppress the magnetic fluctuation. To this end, the Ta impurities are unlikely to be magnetic. We further evaluate the nonmagnetic scattering potential of each Ta impurity. The central mass energy of local density of state for V atom and Ta atom is calculated to be 0.177eV and 0.056eV, respectively. Their difference estimates the scattering potential to be 0.121eV, which is much smaller than the bandwidth on the order of 1eV.

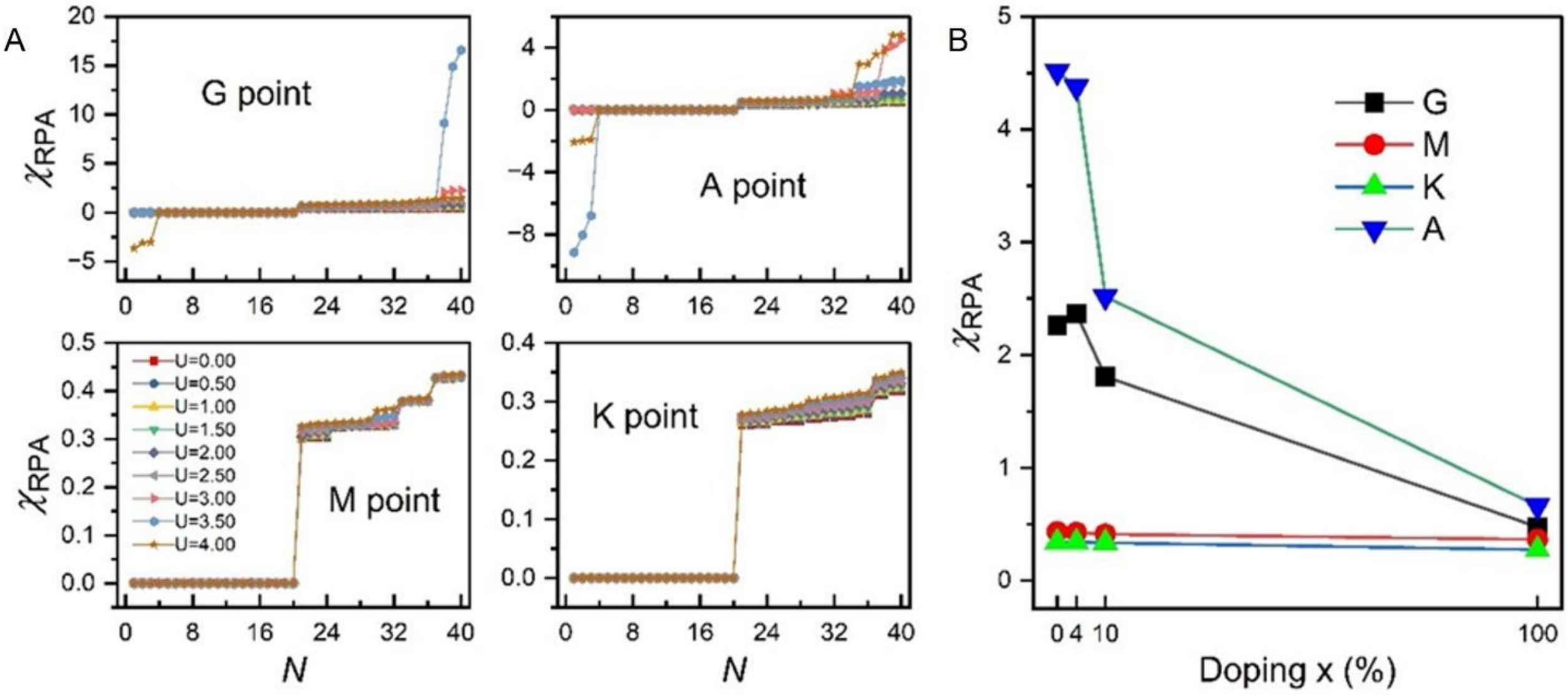

**Fig. S9**. **Calculations supporting the nonmagnetic nature of Ta dopants.** (*A*) Calculated RPA susceptibilities at the high-symmetry points as a function of Hubbard on-site *U* with different doping. (*B*) Leading eigenvalues of the high-symmetry point as a function of doping.

**Real-space Bogoliubov-de Gennes calculations and bond-pairing**

In this section, we study modulated superconducting states in a single-orbital $t - V_1 - V_2$ model on the kagome lattice with a nearest-neighbor (NN) bond-pairing interaction $J$, within a self-consistent real-space Bogoliubov-de Gennes (BdG) approach. Related discussions of modulated states in kagome-lattice systems can be found in Ref. (23, 24). The Hamiltonian is written as:

$$H = H_t + H_V + H_J - \mu \sum\nolimits_{i\sigma} c_{i\sigma}^{\dagger} c_{i\sigma} \quad (1)$$

where

$$H_t = -t \sum\nolimits_{\langle ij\rangle,\sigma} \left( c_{i\sigma}^{\dagger} c_{j\sigma} + \mathrm{H.c.} \right) \quad (2)$$

$$H_V = V_1 \sum\nolimits_{\langle ij\rangle} \hat{n}_i \, \hat{n}_j + V_2 \sum\nolimits_{\langle\langle ij\rangle\rangle} \hat{n}_i \, \hat{n}_j \quad (3)$$

and

$$H_J = -J \sum\nolimits_{\langle ij\rangle,\sigma\tau} \left( \sigma c_{i\sigma}^{\dagger} c_{j\bar{\sigma}}^{\dagger} \right) \left( \tau c_{j\bar{\tau}} c_{i\tau} \right) \quad (4)$$

Here $\langle ij\rangle$ denotes the NN bonds, $\langle\langle ij\rangle\rangle$ denotes the next-nearest-neighbor (NNN) bonds, and $\bar{\sigma} = -\sigma = \pm 1$ and $\bar{\tau} = -\tau = \pm 1$. The NN hopping amplitude $t$ is taken as the energy unit, $V_1$ and $V_2$ are the NN and NNN density interactions, respectively, and $J$ denotes the NN spin-singlet bond-pairing interaction. The density operator is

$$\hat{n}_i = c_{i\uparrow}^{\dagger} c_{i\uparrow} + c_{i\downarrow}^{\dagger} c_{i\downarrow} \quad (5)$$

At the mean-field level, the interaction terms are decoupled into bond and pairing channels. For the bond channel, we introduce the bond field

$$\chi_{ij} = \sum_{\sigma} \langle \, c_{i\sigma}^{\dagger} c_{j\sigma} \rangle \quad (6)$$

which is, in general, complex. The corresponding mean-field contribution is

$$H_\chi = -V_1 \sum\nolimits_{\langle ij\rangle} \left( \chi_{ji} \hat{\chi}_{ij} + H.c. \right) - V_2 \sum\nolimits_{\langle\langle ij\rangle\rangle} \left( \chi_{ji} \hat{\chi}_{ij} + H.c. \right) \quad (7)$$

where

$$\hat{\chi}_{ij} = \sum\nolimits_{\sigma} c_{i\sigma}^{\dagger} \, c_{j\sigma} \quad (8)$$

To describe superconductivity, we consider a nearest-neighbor spin-singlet bond-pairing channel. The corresponding bond-pairing order parameter is defined on each NN bond $\langle ij\rangle$ as

$$\Delta_{ij} = J \langle c_{j\downarrow} c_{i\uparrow} - c_{j\uparrow} c_{i\downarrow} \rangle \quad (9)$$

This order parameter is defined on the bond connecting NN sites $i$ and $j$, rather than on a lattice site. In general, $\Delta_{ij}$ is complex and can be written as

$$\Delta_{ij} = |\Delta_{ij}| e^{i\theta_{ij}} \quad (10)$$

With this definition, the pairing part of the mean-field Hamiltonian is

$$H_\Delta = -\sum\nolimits_{\langle ij\rangle} \left[ \Delta_{ij} \left( c_{i\uparrow}^{\dagger} c_{j\downarrow}^{\dagger} - c_{i\downarrow}^{\dagger} c_{j\uparrow}^{\dagger} \right) + H.c. \right] \quad (11)$$

The full mean-field Hamiltonian is then

$$H_{MF} = H_t + H_\chi + H_\Delta - \mu \sum\nolimits_{i\sigma} c_{i\sigma}^{\dagger} \, c_{i\sigma} \quad (12)$$

where constant terms from the mean-field decoupling are omitted from the BdG matrix.

We solve Eq. (12) self-consistently in real space using the BdG formalism. The fermion operators are expanded as

$$c_{i\sigma} = \sum\nolimits_n \left(u_{i\sigma}^n \gamma_n - \sigma v_{i\sigma}^{n*} \gamma_n^\dagger\right) \tag{13}$$

where $u_{i\sigma}^n$ and $v_{i\sigma}^n$ are the BdG eigenvectors and $E_n$ are the corresponding eigenvalues. The bond field and bond-pairing field are updated self-consistently according to

$$\chi_{ij} = \sum_\sigma \langle c_{i\sigma}^\dagger c_{j\sigma} \rangle \tag{14}$$

and

$$\Delta_{ij} = J\langle c_{j\downarrow} c_{i\uparrow} - c_{j\uparrow} c_{i\downarrow} \rangle \tag{15}$$

The local density is

$$n_i = \sum_\sigma \langle c_{i\sigma}^\dagger c_{i\sigma} \rangle \tag{16}$$

and the chemical potential $\mu$ is determined self-consistently together with $\chi_{ij}$ and $\Delta_{ij}$. In the self-consistent calculation, the filling is fixed at the upper van Hove singularity.

We take $V_1 = 2.0$, $V_2 = 3.0$, and $J = 2.2$, in units of the NN hopping $t$. The self-consistent iteration is performed until a tolerance of $10^{-6}$ is reached. To test the robustness of the solution, we performed calculations starting from multiple independent random initial conditions and with additional small numerical perturbations. In all cases, the self-consistent procedure converged to the same $2 \times 2$-modulated bond-pairing state. The superconducting order parameter has essentially the same amplitude scale in all cases, although the detailed bond configuration may differ microscopically among the converged solutions.

Fig. S10 shows the bond-resolved pairing order in a representative converged state. The left panel displays the NN bond-pairing order parameter plotted directly on the bonds, providing a real-space visualization of the bond-pairing texture. The right panel shows the corresponding magnitude $|\Delta_{ij}|$. In both representations, the modulation is clearly $2 \times 2$, indicating that the bond-pairing state is spatially nonuniform and forms a well-defined superlattice structure in this parameter regime.

Since the superconducting order parameter is defined on bonds rather than on lattice sites, the spatial structure of the state is encoded in the bond-resolved complex field $\Delta_{ij}$. The left panel therefore directly visualizes the real-space distribution of the bond pairing, while the right panel highlights the spatial modulation of its amplitude. Taken together, these results demonstrate that the $2 \times 2$-modulated bond-pairing state is a robust self-consistent solution of the kagome-lattice $t - V_1 - V_2 - J$ model in this parameter regime.

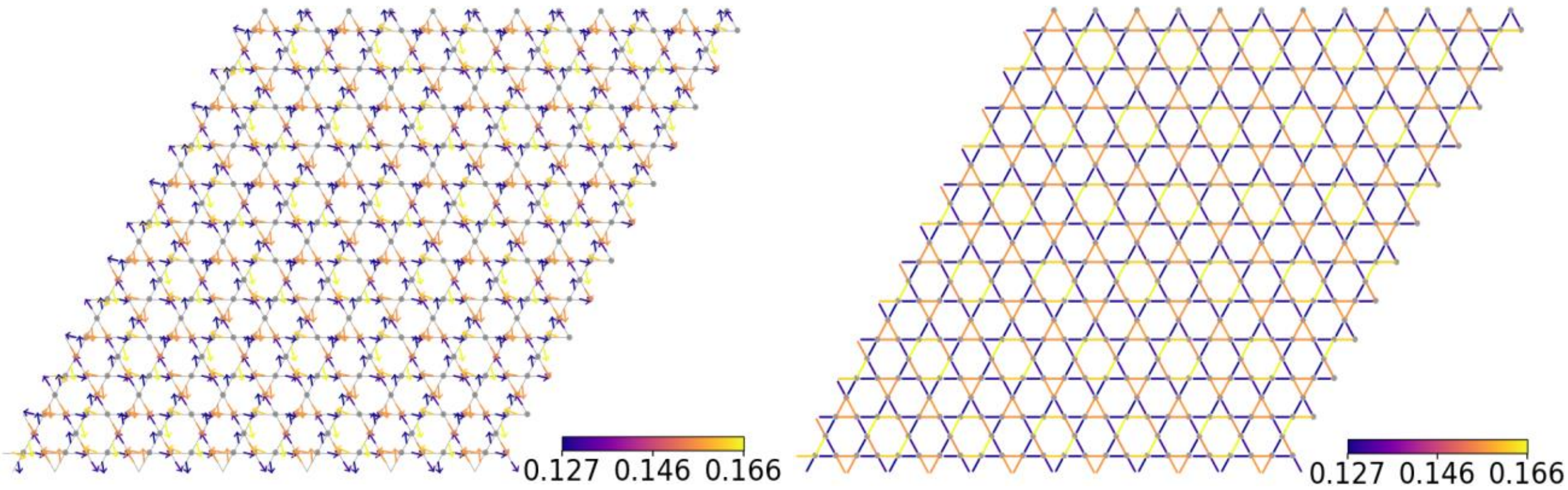

**Fig. S10. Real-space 2×2 PDW in the kagome lattice obtained from the self-consistent BdG calculation.** Left: nearest-neighbor spin-singlet bond pairing displayed in a bond-centered vector representation of the complex order parameter $\Delta_{ij}$. For each bond $ij$, the arrow is placed at the bond midpoint; its length and color scale with $|\Delta_{ij}|$, while its orientation is determined by rotating the bond-direction unit vector $\hat{e}_{ij}$ by the phase $\theta_{ij} = \arg(\Delta_{ij})$. Right: spatial distribution of the pairing magnitude $|\Delta_{ij}|$. Both panels show a pronounced $2 \times 2$ modulation. The same modulated solution is obtained from multiple independent random initial conditions, with nearly identical superconducting order-parameter amplitude.

**Nonmagnetic PDW-breaking effect using Abrikosov-Gor'kov approach**

We specifically focus on the disorder effect in the kagome PDW near the Born limit. Disorder will introduce a random fluctuation in the phase mode, which corresponds to the Goldstone mode arising from the translational symmetry broken. Thus, disorder is unfavorable to PDW states. Based on the minimum self-consistent model (24), we consider 2×2 PDW in the kagome lattice associated with the V-*d* bands, in which the PDW state coexists with a charge order state. At the van Hove singularities, the Hamiltonian holding the intrinsic PDW states can be written as follows in the presence of the bond-attractive interactions:

$$\mathcal{H}_P = -J \sum_{\langle i,j \rangle, \sigma, \tau} \left( \sigma c_{i,\sigma}^{\dagger} c_{j,\bar{\sigma}}^{\dagger} \right) \left( \tau c_{i,\bar{\tau}} c_{j,\tau} \right), \tag{17}$$

where $\langle i,j \rangle$ are nearest bonds connecting sites $i$ and $j$, $\sigma, \tau = -\bar{\sigma}, \bar{\tau} = \pm 1$. In the mean-field approximation, non-vanishing $\sum_{\tau} \langle \tau c_{i,\bar{\tau}} c_{j,\tau} \rangle$ indicates singlet pairing on bond $\langle i,j \rangle$.

In addition, a charge order state can be stabilized by including Coulomb repulsion on neighboring sites with the Hamiltonian form:

$$\mathcal{H}_V = V_1 \sum_{\langle i,j \rangle} n_i n_j + V_2 \sum_{\langle\langle i,j \rangle\rangle} n_i n_j. \tag{18}$$

These interactions can support chiral charge order states, and are compatible with (chiral) PDW states within the charge order state. Assuming that the disorder does not change the bare effective pairing interaction, the leading effect of disorder to suppress the PDW state, is due to the pair breaking effect, similar to that observed in ordinary superconducting states. In order to see the disorder effect clearly, we utilize the Abrikosov-Gor'kov formula (25-27), with the vertex correction up to the Born approximation, as follows:

$$ln\frac{T_c}{\tilde{T}_c} = \chi\left[\psi\left(\frac{1}{2}+x\right) - \psi\left(\frac{1}{2}\right)\right], \tag{19}$$

where $\psi$ is the digamma function, and $\chi = \langle |\Gamma^l(k,q)|^2 \rangle_{\mathrm{FS}} - |\langle \Gamma^l(k) \rangle_{\mathrm{FS}}|^2$ measures the inhomogeneity of the gap function on the Fermi surface.

Our Abrikosov-Gor'kov analysis will be based on the gap function, which is highly anisotropic for the PDW state (24). For simplicity, the gap function can be fitted through the following model function:

$$\Delta(\boldsymbol{k}) = \Delta_0^{\alpha}\, exp(2i\theta_{\boldsymbol{k}}) + \sum_{\theta_p} \Delta_0\, exp(2i\theta_{\boldsymbol{k}}) \left[ e^{-\frac{(\theta_{\boldsymbol{k}}-\theta_p+\delta\theta)^2}{2\Delta\theta^2}} + e^{-\frac{(\theta_{\boldsymbol{k}}-\theta_p-\delta\theta)^2}{2\Delta\theta^2}} \right] \tag{20}$$

where $\theta_{\mathrm{k}}$ is the angular position of the momentum **k** on the Fermi surface, $\Delta_0$ is the amplitude of pairing. In our model function, the double-peak feature is fitted by the Gaussian function located at $\theta_p = 2\pi p/6, p = 0, 1, \ldots, 5$ split by $\pm\delta\theta$ and of width $\Delta\theta$. Moreover, the minigap is included with a power-law suppressed amplitude $\Delta\theta$. Specifically, the parameters here are: $\Delta_0 = 0.03$, $\delta\theta = 0.0375\pi$, $\Delta\theta = 0.0125\pi$, and $\alpha = 3$. Such complex gap function in momentum space mainly results from the phase sign modulation of the PDW in real space (24).

With the relationship shown in Eq. (19), we discuss the disorder effect. First, we consider the isotropic s-wave pairing. Now $\chi = 0$, thus $\tilde{T}_c = T_c$, that is, isotropic s wave pairing is robust against nonmagnetic disorders, which is known as Anderson theorem. Next, we focus on the PDW states as mentioned above. By normalizing the gap function as $\Delta(\boldsymbol{k}) = \tilde{\Delta}_0 \Gamma(k)$ with $\langle |\Gamma^l(k,q)|^2 \rangle_{\mathrm{FS}}=1$, and calculating $\chi = \langle |\Gamma^l(k,q)|^2 \rangle_{\mathrm{FS}} - |\langle \Gamma^l(k) \rangle_{\mathrm{FS}}|^2$, then we obtain the suppression of $T_c$ due to disorder for the kagome PDW, as shown in Fig. 5*F* of the main text.

**Ginzburg-Landau analysis for competing superconducting instabilities**

To understand the enhanced uniform superconductivity and multi-orbital nature of the electron pairing, we further perform a Ginzburg-Landau analysis. From the experimental evidence in our and related studies, the dominant symmetry-broken phases in the system are: 1) a high temperature 2 ×2 charge order $\Delta_{\mathrm{CO}}(\mathbf{M}_i)$, and 2) a phonon driven *s*-wave superconducting order $\Delta_p(\mathbf{\Gamma})$ on the Sb $p_z$-orbital-derived central circular pocket. As the temperature is reduced, V *d*-orbital derived hexagonal pocket develops superconducting pairing with two kinds: 3) a proximity-induced superconducting order on the *d*-orbital pockets, which could be of conventional s-wave type or feature a *d*+i*d* wave gap structure in accordance with predictions from an intrinsically electronic pairing mechanism (28, 29) $\Delta_d(\mathbf{\Gamma})$. However, only the former one can couple to the dominant superconducting order $\Delta_p(\mathbf{\Gamma})$ (30). Hence, we will focus on s-wave pairing on the *d*-orbital state in the following. 4) A PDW state, which arises from an inter-pocket pairing between the *p*- and *d*-pockets $\Delta_{pd}(\mathbf{M}_i)$. This order requires a finite momentum condensate, which is commensurate with the charge order. Therefore, we have four different instabilities with corresponding transition temperatures $T^* \gg T_p > T_{d/pd}$.

Next, we expand the free energy in the Ginzburg-Landau formalism. Since the charge order transition scale is separated from the subsequent superconducting transitions, we do not treat the charge order as a Ginzburg-Landau parameter but incorporate its effects into the symmetry requirements for the free energy functional. In the charge order state, the system exhibits reduced translational symmetry, so all $M_i$ are mapped back to the Γ point. Consequently, we drop the momentum label in the superconducting order parameter and treat them all as ordinary superconducting instabilities. The nature of the PDW phase is thereby encoded in the inter-orbital character of its Cooper pairs.

In the Ginzburg-Landau analysis, the symmetry of the PDW state is highly sensitive to the precise form of the parent charge order parameter (24, 30). However, the nature of the charge-ordered state is still under discussion (23, 31-39). Rather than restricting our analysis to a specific charge order state, we implicitly assume, that the PDW state will transform within an irreducible representation of the enlarged point group $C_{6v}'''$, compatible with the coupling to the irreducible content of the charge order phase. Consequently, we will not explicate all fourth-order terms in the Ginzburg-Landau formalism that would determine the energetically favorable linear combination of multi-dimensional irreducible representations, but instead represent them in a single Ginzburg-Landau coefficient $\beta_i$. For a detailed discussion on the coupling of the different orders on the V sites, we refer to Ref. (30).

To each superconducting order parameter $\Delta_i$, we assign an intrinsic transition temperature $T_c^i$, which contributes to the Ginzburg-Landau functional

$$F_i = \alpha_i |\Delta_i|^2 + \beta_i |\Delta_i|^4 + \mathcal{O}(\Delta_i^6). \tag{21}$$

Near $T_c^i$ the second-order coefficient scales as $\alpha_i \sim \tilde{\alpha}_i (T - T_c^i)$ and $\beta$ is temperature-independent. Since all $T_c^i$ are expected to be sufficiently close, this approximation is valid for investigating the mutual coupling of the SC orders. The full Ginzburg-Landau functional, up to fourth order, is then written as:

$$\begin{gathered} F = \alpha_p |\Delta_p|^2 + \alpha_d |\Delta_d|^2 + \alpha_{dp} |\Delta_{dp}|^2 \\ + a_{01}(\Delta_p \Delta_d^* + h.c.) + a_{02}(\Delta_p \Delta_{pd}^* + h.c.) + a_{12}(\Delta_d \Delta_{pd}^* + h.c.) \\ + \beta p |\Delta_p|^4 + \beta d |\Delta_d|^4 + \beta dp |\Delta_{dp}|^4 \\ + b_{01}(\Delta_p^2 (\Delta_d^*)^2 + h.c.) + b_{02}(\Delta_p^2 (\Delta_{pd}^*)^2 + h.c.) + b_{12}(\Delta_d^2 (\Delta_{pd}^*)^2 + h.c.) \\ + b_{012}\left(\Delta_p \Delta_d^* |\Delta_{pd}|^2 + h.c.\right) + b_{201}(\Delta_{pd} \Delta_p^* |\Delta_d|^2 + h.c.) \\ + b_{120}\left(\Delta_d \Delta_{pd}^* |\Delta_p|^2 + h.c.\right) + \mathcal{O}(\Delta_i^6). \end{gathered} \tag{22}$$

The additional second-order terms arise from the coupling between different order parameters. The presence of $a_{02}$, $a_{12} \propto \Delta_{\mathrm{CO}}$ is a direct result of the underlying charge order phase, as it stems from the band-folding inside the reduced Brillouin zone of the 2 ×2 state. The actual superconducting transition temperature is determined as the temperature at which the second-order coefficient matrix becomes singular, *i.e.*

$$\det\begin{pmatrix} \alpha_p & a_{01} & a_{02} \\ a_{01} & \alpha_d & a_{12} \\ a_{02} & a_{12} & \alpha_{pd} \end{pmatrix} = 0. \qquad (23)$$

Since the coupling between the order parameters is already captured by the second-order terms, which are dominant near the phase transition, we focus on these terms and disregard the fourth-order terms in the following analysis.

In the pristine case, the resulting superconducting phase on the $d$-pocket is exclusively governed by $\Delta_{pd}$, leading to a partially gapped Fermi surface with residual Fermi arcs (2, 40). This can be understood within the Ginzburg-Landau picture: The second-order coefficients in Eq. (22) are determined by the overlaps between the different SC orders with the bare particle-particle kernel

$$a_{ij} = \sum_{\{o_i\}\boldsymbol{k}} \int d\tau \left\langle \Delta_i^{o_1 o_2}(\boldsymbol{k}) \middle| G_{o_1 o_3}(\boldsymbol{k},\tau) G_{o_2 o_4}(-\boldsymbol{k},\tau) \middle| \Delta_j^{o_3 o_4}(\boldsymbol{k}) \right\rangle. \qquad (24)$$

where $G_{o_1 o_2}(\boldsymbol{k},\tau)$ is the single-particle Green's function in the charge-ordered phase. For all possible coupling terms, the Green's functions only contribute non-zero terms at the intersection of the backfolded pockets, where both $\Delta_{pd}$ and $\Delta_d$ can be of comparable size. While $\Delta_d$ may also exhibit a finite gap away from the band crossings to gain condensation energy, *i.e.* $T_c^d > T_c^{pd}$, the orbital structure of the gap functions suggests $a_{02} > a_{01}$ due to the more favorable orbital overlap between the superconducting states. With $T_c^p \gg T_c^d$, $T_c^{pd}$, the competition of these two effects determines the superconducting ground state. In the pristine case, the effect of $a_{02}$ seems to dominate over the intrinsic transition temperature difference.

In the system with randomly distributed Ta dopants, two striking effects can be observed: i) the $d$-orbital pockets are substantially gapped, reducing the remaining density of states within the superconducting gap; ii) the superconducting transition temperature is substantially enhanced compared to the pristine case. Since the suppression of the charge order is marginal and observed at a temperature well above the superconducting transition, we do not account for changes in the charge order-mediated coupling in the Ginzburg-Landau functional of Eq. (22). It is known that disorder can lead to an enhancement of $T_c$ if impurities alter the density of states distribution in a way that increases the condensation energy (41). We incorporate the effect of disorder into our phenomenological theory by accounting for its influence on the second-order coefficients. While the s-wave superconducting orders $\Delta_p$ and $\Delta_d$ are robust against non-magnetic impurities, $\Delta_{pd}$ suffers from a strong suppression of $T_c$ in the presence of disorder (as evidenced by the Abrikosov-Gor'kov calculations). We note that the composite $pd$-nature of the superconducting state is crucial: it has been shown that the sublattice polarization on the $d$-orbital derived Fermi surface protects also unconventional superconducting orders from local disorder (42). However, disorder destroys the phase coherence of the inter-pocket Cooper pairs, leading to a substantial decrease of $T_c^{pd}$. We mimic this effect in Eq. (23) by decreasing $T_c^{pd}$ and simultaneously increase $T_c^p$ to account for the enhanced electron phonon coupling for conventional pairing potential on the Sb $p$-orbital derived pocket.

Additionally, impurities significantly increase the inter-pocket scattering (43), which we incorporate by a change of $a_{01}$ while keeping the other parameters constant. Fig. S11 shows the qualitative evolution of the gap structure as a function of the varied parameter. We observe a crossover from $\Delta_{pd}$ to $\Delta_d$ descendant superconducting instability on the *d*-orbital pocket, which coincides with quasi-particle interference measurements. Similarly, the overall superconductivity is enhanced for a sufficiently large inter-pocket coupling beyond the increase from the electron phonon coupling incorporated in the Ginzburg-Landau parameter: it exceeds the increase of $T_c^p$ by a factor of 2 for the chosen parameters.

While our phenomenological Ginzburg-Landau theory does not allow for a quantitative estimate of the $T_c$ enhancement due to Ta dopants, it clearly showcases that the multi-orbital and multi-pocket nature with different competing and cooperating superconducting orders in the non-trivial Fermi surface topology allows to considerably enhance $T_c$ beyond the 30% increase predicted by conventional intra-pocket s-wave pairing mechanism.

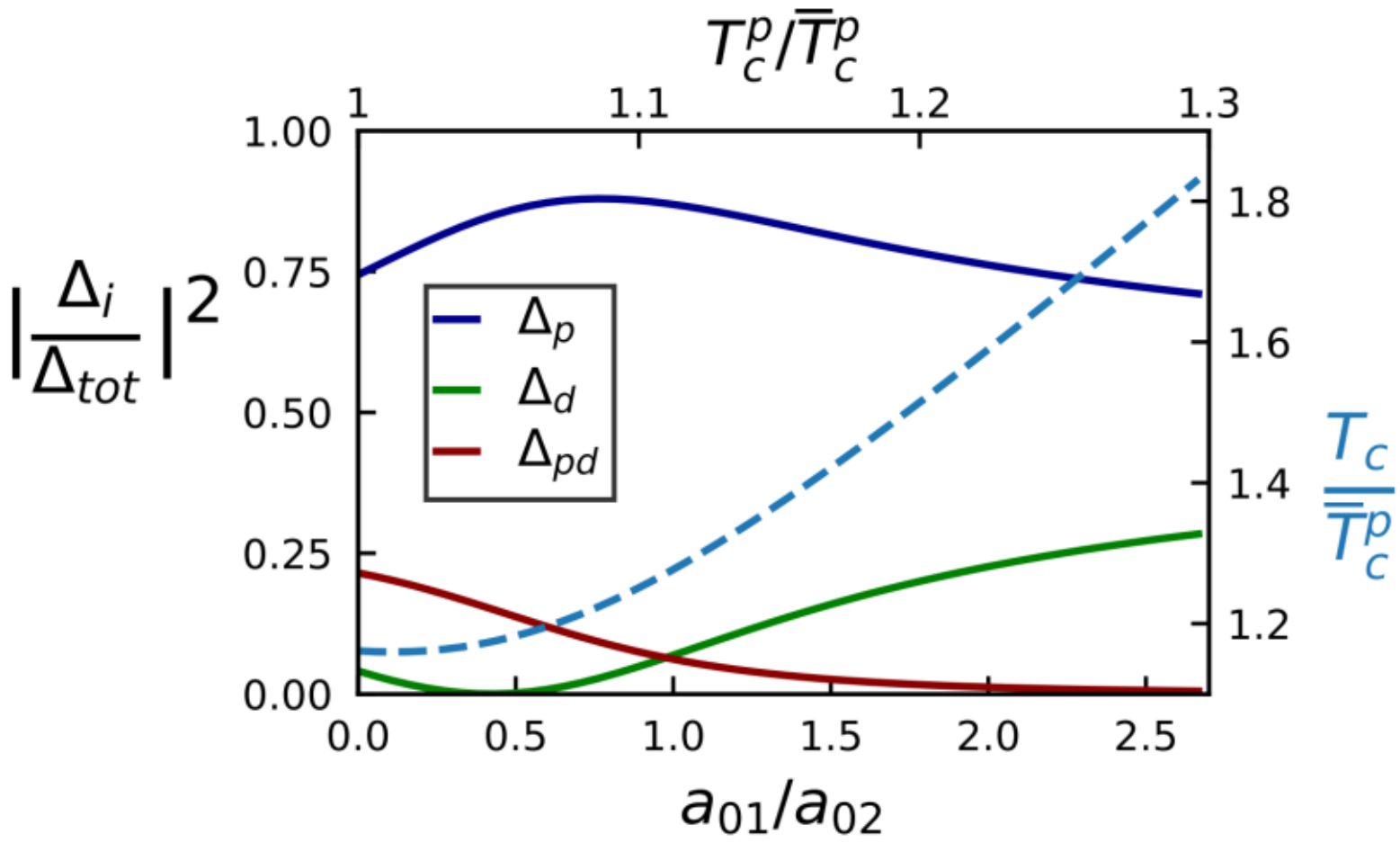

**Fig. S11. Ginzburg-Landau analysis for competing superconducting instabilities.** Schematic temperature evolution upon introduction of non-magnetic impurities. We have used $\alpha_d = \alpha_{pd} = 0.8\alpha_p$ and fixed $T_c^d = 0.3$ and $a_{02} = a_{12} = 0.3$. The effect of Ta dopants is mimicked by a 30% increase of $T_c^p$ compared to the pristine value $\bar{T}_c^p = 1$ as indicated by the electron-phonon coupling calculations. The simultaneous suppression of $T_c^{pd}$ due to nonmagnetic PDW breaking effect is modelled by $T_c^{pd}$=0.3(2-$T_c^p/\bar{T}_c^p$). The qualitative behavior of the gap and $T_C$ evolution is independent of the chosen parameters. Only the crossover between $\Delta_{pd}$ and $\Delta_d$ dominated regimes is slightly shifted.

## SI References